\documentclass[journal]{IEEEtran}
\usepackage{textcomp}

\usepackage{url}
\usepackage{verbatim}
\usepackage{graphicx}
\def\BibTeX{{\rm B\kern-.05em{\sc i\kern-.025em b}\kern-.08em
    T\kern-.1667em\lower.7ex\hbox{E}\kern-.125emX}}
\usepackage{balance}

\usepackage{colortbl}  
\usepackage{array}   
\usepackage{cite}
\usepackage{mathrsfs}            
\usepackage{amsmath,amssymb,amsfonts}
\usepackage{multirow}

\usepackage{algorithm} 
\usepackage{algorithmic}

\usepackage{enumerate}
\usepackage{enumitem} 
\usepackage{mathrsfs}            
\usepackage{multirow}
\usepackage{enumerate}
\usepackage{booktabs}
\usepackage{threeparttable}
\usepackage{subfigure}
\usepackage{tablefootnote}
\usepackage{listings}
\usepackage{hyperref}
\usepackage{bbding}
\usepackage{balance}
\usepackage{hhline} 

\usepackage{xpatch}
\usepackage{xcolor}
\usepackage{array}
\usepackage{verbatimbox}

\begin{document}

\title{From Exposure to Internalization: Dual-Stream Calibration for In-context Clinical Reasoning}

\author{Chuang~Zhao, Hongke Zhao, Xiaofang Zhou,~\IEEEmembership{Fellow, IEEE}, Xiaomeng~Li,~\IEEEmembership{Senior Member, IEEE}
        
\IEEEcompsocitemizethanks{
\IEEEcompsocthanksitem C. Zhao and X. Li are with the Department of Electronic and Computer Engineering, The Hong Kong University of Science and Technology, Hong Kong, SAR, China;  (e-mail: czhaobo@connect.ust.hk, eexmli@ust.hk).  X. Li is the corresponding author.
\IEEEcompsocthanksitem H. Zhao is with the College of Management and Economics, Laboratory of Computation and Analytics of Complex Management Systems (CACMS), Tianjin University, Tianjin 30072, China; (e-mail: hongke@tju.edu.cn) 
\IEEEcompsocthanksitem X. Zhou is with the Department of Computer Science and Engineering, The Hong Kong University of Science and Technology, Hong Kong, SAR, China;  (e-mail: zxf@ust.hk). 
}}

\markboth{XXXXXXXXXX}%
{Shell \MakeLowercase{\textit{et al.}}: Bare Advanced Demo of IEEEtran.cls for IEEE Computer Society Journals}

\maketitle

\begin{abstract}
Contextual clinical reasoning demands robust inference grounded in complex, heterogeneous clinical records. While state-of-the-art fine-tuning, in-context learning (ICL), and retrieval-augmented generation (RAG) enable knowledge exposure, they often fall short of genuine contextual internalization: dynamically adjusting a model’s internal representations to the subtle nuances of individual cases at inference time.
 To address this, we propose Dual-Stream Calibration (DSC), a test-time training framework that transcends superficial knowledge exposure to achieve deep  internalization during inference.
DSC facilitates input internalization by synergistically aligning two calibration streams. Unlike passive context exposure, the Semantic Calibration Stream enforces a deliberative reflection on core evidence, internalizing semantic anchors by minimizing entropy to stabilize generative trajectories. Simultaneously, the Structural Calibration Stream assimilates latent inferential dependencies through an iterative meta-learning objective. By training on specialized support sets at test-time, this stream enables the model to bridge the gap between external evidence and internal logic, synthesizing fragmented data into a coherent response.
Our approach shifts the reasoning paradigm from passive attention-based matching to an active refinement of the latent inferential space. Validated against thirteen clinical datasets, DSC demonstrates superiority across three distinct task paradigms, consistently outstripping state-of-the-art baselines ranging from training-dependent models to test-time learning frameworks.

\end{abstract}





\section{Introduction}\label{sec:intro}
\IEEEPARstart{C}{linical} reasoning stands as a unique knowledge-intensive frontier,  distinct from conventional linguistic comprehension~\cite{naturewang2025medical,naturesinghal2025toward}. 
Unlike standard question answering (QA), this task mandates that Large Language Models (LLMs) synthesize intricate, multi-faceted clinical records. These records include comparable patient histories, longitudinal laboratory results, and diagnostic narratives, enabling models to conduct rigorous causal inference and draw accurate, well-justified conclusions~\cite{datadiag,zhaokdd,dwu2024survey}. The central obstacle lies in enabling these models to effectively internalize and harness the input evidence during inference, moving beyond highly speculative leaps or stochastic guessing.

\begin{figure}[!ht] 
\centering
\subfigure[Internalization.]{ 
\begin{minipage}[t]{0.325\linewidth}
\centering
\includegraphics[width=\linewidth,height=0.95\linewidth]{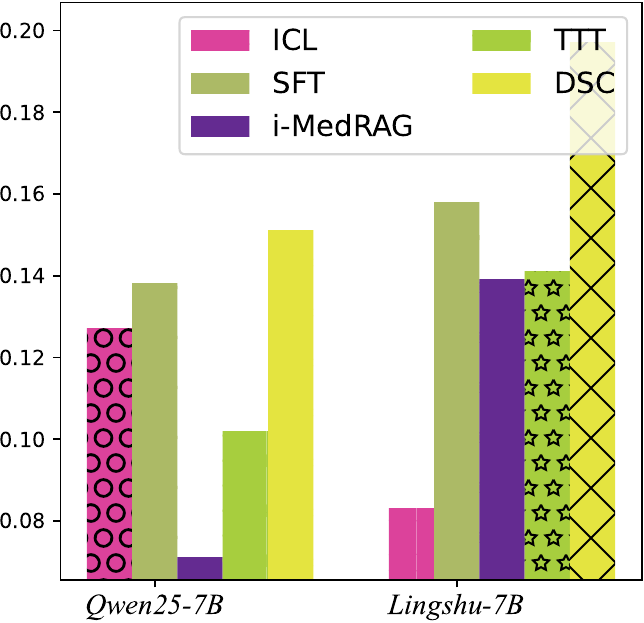}
\label{fig:motiv:int}
\end{minipage}%
}%
\subfigure[Generation Entropy.]{
\begin{minipage}[t]{0.325\linewidth}
\centering
\includegraphics[width=\linewidth,height=0.95\linewidth]{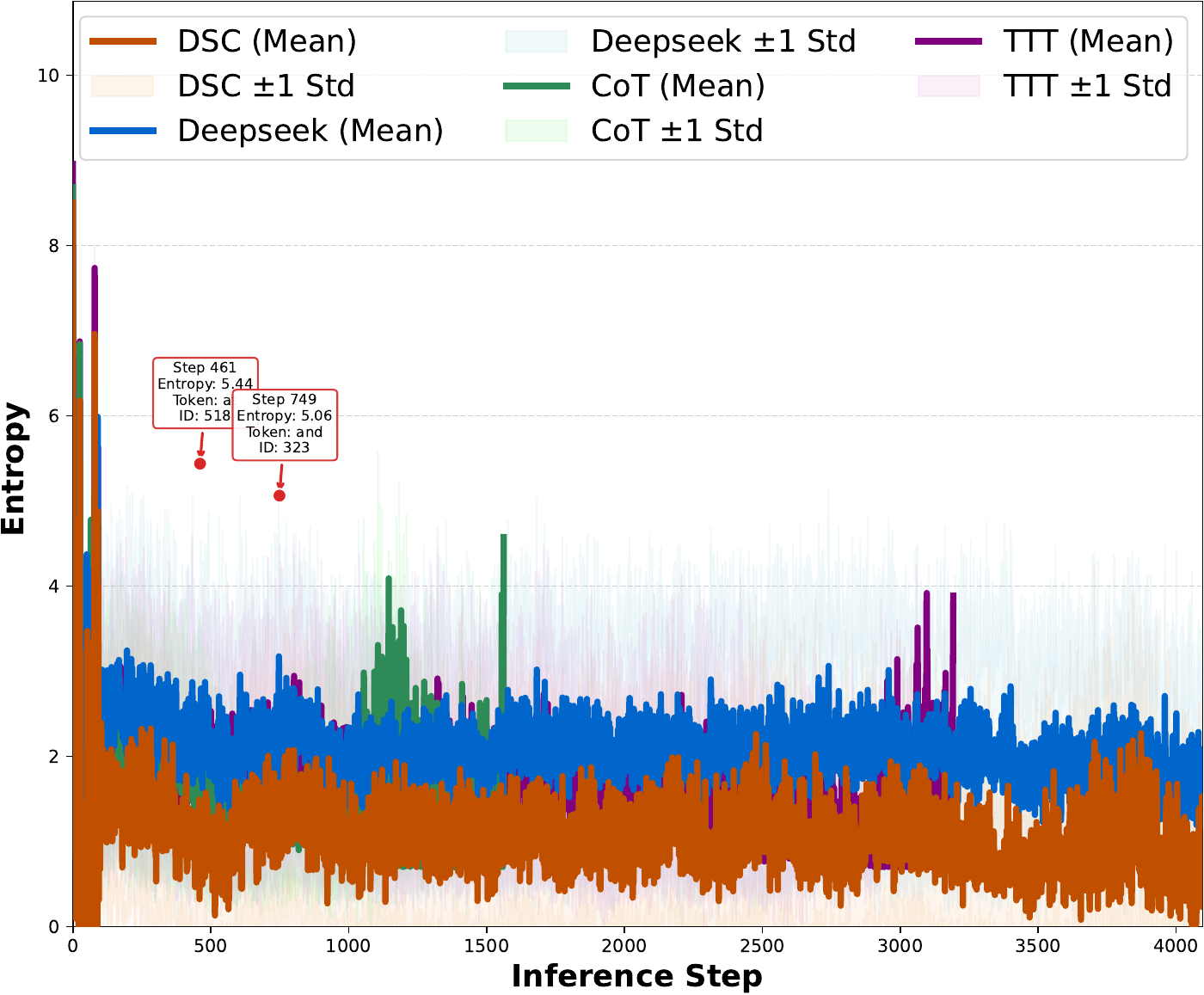}
\label{fig:motiv:ent}
\end{minipage}%
}%
\subfigure[In-context Disturb.]{ 
\begin{minipage}[t]{0.325\linewidth}
\centering
\includegraphics[width=\linewidth,height=0.95\linewidth]{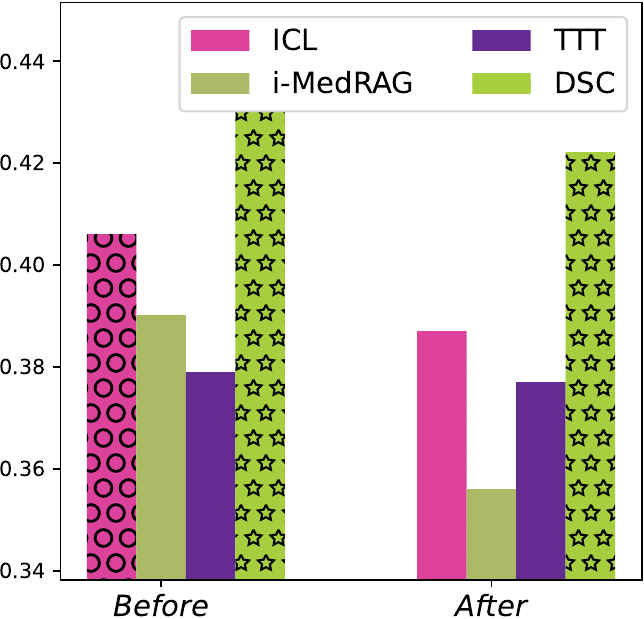}
\label{fig:motiv:con}
\end{minipage}%
}%
\centering
\setlength{\abovecaptionskip}{-0.15cm}   
\setlength{\belowcaptionskip}{-0.1cm}   
\caption{Motivations. Fig.~\ref{fig:motiv:int} presents the ROUGE-L scores of various models evaluated on auxiliary QA pairs per sample. These QA pairs are generated from the eLife dataset's context using DeepSeek-V3~\cite{deepseekr1} and subsequently provided as input for model evaluation. Qwen~\cite{qwen25} and Lingshu~\cite{lingshu} denote different LLM backbones. Fig.~\ref{fig:motiv:ent} depicts the predictive entropy fluctuations during the generative process. Fig.~\ref{fig:motiv:con} evaluates model resilience (ROUGE-L) to contextual order on the eLife dataset, where the order of in-context demonstrations is randomly perturbed.}
\label{fig:motiv}
\vspace{-0.1cm}
\end{figure}

A critical review of existing methodologies reveals limitations across two primary paradigms. The training-based paradigm, represented by Supervised Fine-Tuning (SFT)~\cite{sft,survey-sft2} and Reinforcement Learning (RL)~\cite{grpo,survey-rlhf}, attempts to fossilize clinical expertise within the model's parametric weights during training. 
While enhancing domain familiarity, this strategy induces a frozen reasoning logic that is ill-equipped for the fluid nature of medicine, where the prohibitive cost of large-scale retraining renders models lag behind evolving clinical guidelines.
More critically, this rigid parametric dependency creates a generalization bottleneck: when faced with Out-of-Distribution (OOD) scenarios~\cite{zxu2025large,survey-tts2} where patient symptoms deviate from training templates, the model's internal logic often collapses, resulting in brittle performance in high-stakes clinical environments. To transcend these costs and static constraints, recent research has shifted toward independent adaptation at the inference stage. The test-time tuning-free paradigm achieves this through dynamic evidence synthesis, including medical context-driven methods like RAG~\cite{lightrag,ColaCare}, ICL~\cite{ICL,Liqingtkde}, and scaling-driven strategies like Chain-of-Thought (CoT)~\cite{cot} and multi-agent collaboration~\cite{MDAgents,MedAgents}.
While promising, these approaches fall into a ``passive observation" trap, treating context and answer as sequential pattern-matching problem rather than an explicit derivation. They lack a mechanism for dynamic path calibration, failing to interrogate whether the input evidence genuinely supports the diagnostic claim. Consequently, the model’s output remains an educated guess rather than a structured derivation; by failing to actively recalibrate its response against the evidence's causal weight, the model merely ``sees" rather than ``comprehends" the context. This leads to increased output uncertainty and sensitivity to context perturbations, as illustrated in Fig.~\ref{fig:motiv}.

Most recently, the test-time tuning-based paradigm represented by TTT~\cite{TTT2025} and SLOT~\cite{slot} has emerged as a frontier for inference-time refinement. However, these methods exhibit two specific limitations that reduce their effectiveness in clinical reasoning.
First, these methods apply uniform optimization weight across all input tokens. In clinical records where administrative notes, similar patient histories, and other auxiliary text substantially outnumber diagnostically relevant query, this uniform weighting causes the model to allocate update capacity to low-signal tokens. Accordingly, instead of sharpening its reasoning, the model dilutes its specialized medical knowledge by overfitting to the entropy of auxiliary records, a failure mode evidenced by the performance degradation in OOD settings in Fig.~\ref{fig:ood} and heightened generation entropy in Fig.~\ref{fig:motiv:ent}.
Second, these paradigms treat complex clinical backgrounds as flat token sequences, which does not capture the structured dependencies among longitudinal observations. For instance, diagnosing a patient with an atypical presentation of sepsis may require reasoning over structurally similar prior cases, identifying that a comparable pattern of inflammatory markers and vital sign trajectories in a previous patient resolved to the same diagnosis. Without a mechanism to leverage such inter-case inferential paths, the updated parameters cannot transfer relational evidence from comparable patients to the target case, limiting the model's ability to resolve ambiguous diagnostic targets from heterogeneous clinical observations. This limitation is reflected in the high context dependence and poor internalization  observed in Fig.~\ref{fig:motiv:con} and Fig.~\ref{fig:case:int}.

Motivated by the shared limitations of both test-time tuning paradigms and the need to move beyond superficial knowledge exposure, we propose \textbf{D}ual-\textbf{S}tream \textbf{C}alibration (DSC) for in-context clinical reasoning. 
Our central technical insight posits that achieving true contextual understanding requires a targeted, two-pronged calibration of the input representation at inference time, thereby shifting the LLM from a passive observer to an active agent of deep internalization. We commit to rectifying the input's deficiencies from both the semantic and structural perspectives. 
Specifically, the Semantic Calibration Stream reduces spurious uncertainty through a dynamic entropy detection and revision strategy operating as a self-reflective loop. Unlike indiscriminate optimization, our model selectively differentiates between context and query, monitoring the generation process in long-short windows to surgically revise high-entropy tokens. This ensures that every diagnostic claim represents a converged state of high-confidence clinical evidence, effectively grounding the model's generative logic in verifiable certainty, as depicted in Fig.~\ref{fig:case:ent}. Concurrently, the Structure Calibration Stream leverages a meta-learning paradigm to redefine the model’s interaction with contextual information. Rather than treating the clinical background as a static reference, this stream employs an alternating framework of specialized support sets and meta-queries to train the model on the act of inference itself. In practice, this mechanism empowers the LLM to master the inferential protocols of context-to-answer mapping. By treating the input as a  navigable knowledge base, the model learns to reconstruct the structural pathways that link analogous patient topographies to tailored diagnostic conclusions. This transforms the reasoning process from a black-box extrapolation into a structural derivation, ensuring that every conclusion is a direct byproduct of navigated evidence synthesis, as evidenced in Fig.~\ref{fig:motiv:con}.

To summarize, the contributions of this work are threefold:
\begin{itemize}[leftmargin=12pt]
    \item We propose DSC, a novel test-time training framework that  shifts the LLM paradigm from superficial context exposure to deep context internalization, enabling adaptive and accurate clinical reasoning at inference time. 
    \item We develop two specialized, fine-grained calibration strategies: the dynamic entropy detection and elimination for semantic calibration, and an iterative structural calibration for reconstructing context-answer dependencies. 
    \item We conduct extensive experiments on 13 challenging clinical benchmarks, demonstrating that our DSC framework consistently establishes new state-of-the-art performance, validating its superior effectiveness and efficiency. 
\end{itemize}

\section{Related Work}\label{sec:rel}
\begin{figure}[!tp]
    \centering
    \setlength{\abovecaptionskip}{0cm}
    \setlength{\belowcaptionskip}{0cm}
\includegraphics[width=0.95\linewidth,height=0.35\linewidth]{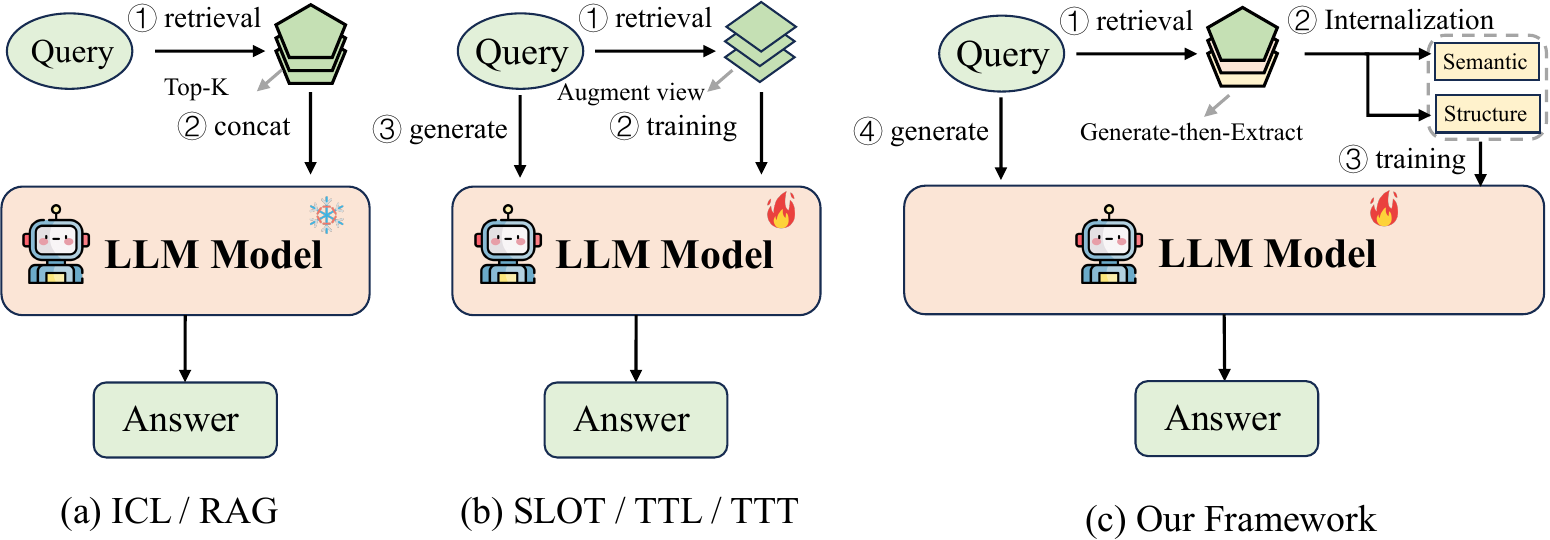}
    \caption{Key difference between ICL, RAG, TTL-based baselines, and the proposed DSC. Our core innovation introduces a dual-calibration framework that enforces semantic understanding and structure robustness, transforming passive information retrieval into active contextual internalization.}
    \label{fig:diff}
\end{figure}

We review the closely related work, highlighting both connections and distinctions. For clarity, we present the key difference with the closely related work in Fig.~\ref{fig:diff}.

\subsection{Clinical Reasoning}\label{sec:rel:health}

Clinical reasoning transcends superficial pattern matching, requiring high-stakes inferential synthesis over complex and often noisy longitudinal patient records~\cite{zyinhao2025,zhaowww,zliutkde}. Unlike traditional domain-based question answering, it necessitates the active construction of diagnostic trajectories from fragmented evidence rather than simple factual lookups~\cite{zhaotkde,zhaotois}.

The evolution of clinical LLM adaptation is defined by a shift from static training alignment toward dynamic, inference-time reasoning. The training-based paradigm, represented by SFT and RL, seeks to fossilize clinical expertise within parametric weights~\cite{Magic,DiaRL}. While effective for domain alignment, these high-cost methods yield rigid models prone to brittle collapse in OOD scenarios, as they lack the fluid adaptability required for evolving patient cases. To circumvent the costs of retraining, the test-time paradigm encompassing RAG, ICL, and Multi-agent workflows shifts toward dynamic knowledge utilization by augmenting prompts with external evidence or collaborative scaling~\cite{zXiong2025Improving, MedAgents}. While frameworks like AgentSimp~\cite{AgentSimp}, TAGS~\cite{tags}, and ColaCare~\cite{ColaCare,ColaCare2} significantly enhance task-level performance through diverse agent roles and enriched data sourcing, they remain tethered to an extrinsic optimization logic. By focusing on the orchestration of external workflows rather than the intrinsic calibration of the model's latent states, these methods expand the search space without refining the inferential precision of the underlying LLM. Consequently, the model acts as a passive aggregator of information, failing to resolve the core tension between dense contextual noise and the causal clarity required for diagnostic certainty.

\textit{Moving beyond passive knowledge exposure, DSC achieves active logical internalization by synchronizing semantic grounding with structural navigation. While prior test-time methods in medical treat clinical records as static, uncalibrated inputs, DSC transforms the context into a dynamic inferential space. By resolving spurious uncertainty through its semantic stream and bridging the context-to-answer gap through its structural stream, DSC ensures that generative logic is no longer a black-box extrapolation, but a rigorous, evidence-anchored derivation specifically engineered for high-stakes clinical reasoning.}

\subsection{Test-time Scaling}\label{sec:rel:tes}
Test-time scaling (TTS) refers to a class of techniques designed to enhance a model's reasoning capabilities by strategically increasing computational expenditure during the inference phase~\cite{zxihongdual,TTL,sgen}. Rather than relying solely on a fixed-cost forward pass, test-time scaling seeks to elicit higher-order intelligence from pre-trained weights through mechanisms such as search-based reasoning, iterative refinement, or local parameter adaptation~\cite{survey-tts}.

Existing test-time scaling methodologies can be broadly categorized into prompt-level expansion and parameter-level optimization~\cite{survey-tts2,survey-tts}.
Prompt-level expansion techniques dynamically adjust inference processes without modifying model parameters, focusing on guiding the model’s reasoning through structured interactions. CoT prompting~\cite{cot} is a foundational approach that elicits step-by-step reasoning via explicit instructions, with its variants like Tree-of-Thoughts~\cite{treecot} extending this to branching reasoning paths for complex problem exploration. Multi-agent collaboration, exemplified by MedAgents~\cite{MedAgents} and MDAgents~\cite{MDAgents}, further enhances scaling by coordinating multiple model instances to generate diverse solutions or verify each other’s outputs, harnessing collective reasoning to improve reliability. 
Parameter-level optimization refines the model’s internal parameters to inherently enhance test-time reasoning, reducing reliance on external prompting~\cite{TTT2025}. For example, Test-time Learning-based (TTL) baselines including TLM~\cite{TTL}, TTT~\cite{TTT2025}, and SLOT~\cite{slot}, optimize model parameters to lower perplexity toward input queries, enabling more stable and focused reasoning by aligning the model’s internal representations with task-specific inputs.  Yet, current implementations suffer from a critical objective-task mismatch. This indiscriminate optimization not only risks corrupting the model's linguistic coherence but also fails to capture the latent structural dependencies within medical records, leaving the model prone to hallucinating within high-uncertainty voids.

\textit{Unlike prompt-level expansion techniques, such as CoT or Multi-agent collaboration, DSC deepens the internalization of task-specific evidence without relying on fragile prompt engineering or incurring the prohibitive token overhead of multi-agent architectures. Furthermore, unlike existing parameter-level optimization methods like TLM and SLOT, which suffer from a notable objective-task mismatch by optimizing all tokens indiscriminately, DSC implements a context-query separated, long-short windows detection and optimization. By simultaneously purging semantic uncertainty through entropy-driven calibration and reconstructing latent inferential dependencies via meta-learning, DSC achieves an output-level grounding absent in prior works. This represents a paradigm shift from simple context inclusion to deep, structure-aware context manipulation, ensuring robust reasoning fidelity in high-stakes clinical scenarios. }

\section{Proposed Method}\label{sec:pro}
We first introduce the preliminaries, then provide an overview of the proposed DSC, and detail submodules.
\begin{table}\small
\centering
\setlength{\abovecaptionskip}{-0.05cm}
\setlength{\belowcaptionskip}{-0.1cm} 
\caption{Comprehensive Mathematical Notations.}
\label{tab:math}
\resizebox{0.48\textwidth}{!}{
\begin{tabular}{c|l}
\toprule
\textbf{Notations}& \textbf{Descriptions} \\
\hline
$\mathcal{M}(\theta)$ & Frozen Large Language Model  \\
$\mathcal{C}, \mathcal{Q}, \mathcal{A}$ & Retrieved Context, Input Query, Answer \\
$\mathbf{X}$ & Total Input \\
$\mathbf{H}_\mathcal{C}, \mathbf{H}_\mathcal{Q}$ & Token Embeddings of $\mathcal{C}$ and $\mathcal{Q}$ \\
$\boldsymbol{\delta}_\text{sem}, \boldsymbol{\delta}_\text{str}$ & Semantic/Structure Calibration Vectors (Trainable) \\
$m, n, k$ & Sequence Lengths of $\mathcal{C}$, $\mathcal{Q}$, and $\mathcal{A}$ \\
$\mathbf{H}^*$ & Calibrated Embeddings \\
\hline
$H(p_i)$ & Predictive Entropy at Context Token $i$\\
$N_{\text{short}}$ & Short Window Length \\
${y}_{i}$ & $i$-th Generated Token  \\
$\mu_{\text{s}}, \mu_{\text{l}}$ & Local and Global (Long) Entropy Averages \\
$\tau$ & Dynamic Threshold Multiplier for Uncertain Tokens \\
$\mathcal{U} $ & Set of Uncertain (High-Entropy) Generated Tokens\\
$\mathcal{I}$ & Prompt Instruction \\
$\mathcal{L}_{\text{ent}}$ & Adaptive Entropy Minimization Loss \\
$\mathcal{L}_{\text{rcf}}$ & Recalibration Factor Loss\\
\hline
$M$ & Number of Context Permutations \\
$\text{Context}^{l}_{i}$ & $l$-th Permuted Context of $\mathrm{x}_{i}$ \\
$(\mathcal{P}_\text{meta}^{i,l},\mathrm{y}_{i})$ & $l$-th Meta-training Input, Answer of $\mathrm{x}_{i}$  \\
$\mathcal{L}_{\text{str}}$ & Structural Calibration Stream Loss \\
$\gamma$ & Trade-off weights for $\mathcal{L}_{\text{sem}}$ and $ \mathcal{L}_{\text{str}}$\\
$\mathcal{L}_{\text{dsc}}$ & Total Dual-Stream Calibration Loss \\
\bottomrule
\end{tabular}}
\vspace{-1em}
\end{table}

\subsection{Preliminaries}\label{sec:med:pre}

\noindent\textbf{Clinical Dataset:} Unlike conventional general-domain QA that relies on straightforward factoid retrieval, the datasets in this study feature sophisticated clinical narratives that demand high-order, multi-hop reasoning. Each instance consists of a corresponding clinical query $\mathcal{Q}$,  the background $\mathcal{B}$ (e.g., patient-doctor dialogue, patient history, lab results), and the ground truth answer $\mathcal{A}$. Consistent with~\cite{ColaCare,TTT2025}, we integrate the background information $\mathcal{B}$ and the query $\mathcal{Q}$ into a consolidated query to ensure a comprehensive contextual representation. The final query contains multiple tokens, i.e., $\mathcal{Q} = \{q_1, q_2, \dots, q_n\}$. 

\noindent\textbf{Task Formulation:} Our goal is to train a framework, specifically the DSC framework, that enhances a pre-trained Large Language Model, denoted as $\mathcal{M}(\cdot;\theta)$, to perform {in-context clinical reasoning} through the systematic synthesis of evidence from similar patient profiles. Given a context $\mathcal{C}$ (comprising recalled $K$ demonstrations) and a query $\mathcal{Q}$, the objective is for the model to generate the correct answer $\mathcal{A}$ by effectively internalizing the contextual evidence.
Following the {In-context learning} paradigm~\cite{ICL,survey-icl}, the initial input to the LLM is constructed by concatenating the query and the context:
\begin{equation}\label{eq:1}
\mathbf{X} = [\mathcal{C}; \mathcal{Q}],
\end{equation}
where $[\cdot; \cdot]$ denotes the concatenation operation. Initially, the frozen LLM $\mathcal{M}$ processes the total input $\mathbf{X}$ to generate a final hidden representation $\mathbf{H} \in \mathbb{R}^d$ at the penultimate layer. The model's classification or token generation mechanism is then defined by a linear transformation followed by a softmax activation function:
\begin{equation}\label{eq:2}
P(\mathcal{A} | \mathbf{X}) = \text{Softmax}(W_\text{logits} \cdot \mathbf{H}), 
\end{equation}
where $W_\text{logits}$ denotes the weights of the pre-trained language modeling head.
In this vanilla setting, the probability $P$ of the answer $\mathcal{A}$ is formulated as:
\begin{equation}\label{eq:3}
P(a_1, \dots, a_k | c_1, \dots, c_m, q_1, \dots, q_n) = \prod_{i=1}^{k} P(a_i | a_{<i}, \mathcal{C}, \mathcal{Q}),
\end{equation}
where $\mathcal{C} = \{c_1, c_2, \dots, c_m\}$ typically represents a sequence of tokens. Crucially, we assume that the query and context $\mathcal{C}$ are rich with latent information. As such, the reasoning phase must move beyond simple exposure to achieve precise extraction and internalization of these underlying nuances. To mitigate this, our framework introduces two lightweight adaptations $\delta$. By perturbing the representation into $\mathbf{H}^{*} = \mathbf{H} + \delta_{\text{sem}} +\delta_{\text{str}}$, we effectively recalibrate the inputs to the $W_\text{logits}$ head. This transformation acts as an output-level adapter, designed to crystallize the input evidence and maximize the reasoning fidelity of the frozen model $\mathcal{M}$.
 Consequently, the modified conditional probability is formulated as,
 \begin{equation}\label{eq:4}
P(\mathcal{A} | \mathbf{X}; \delta) = \text{Softmax}(W_\text{logits} \cdot \mathbf{H}^{*}).
 \end{equation}
 To guide the optimization of $\delta$, we design specialized semantic and structural regularized objectives in Section~\ref{sec:med} that enforce consistency during the inference process.
By shifting the latent vector precisely before the logit layer, the framework effectively recalibrates the model's focus, acting as a surgical intervention to maximize reasoning fidelity while keeping the extensive knowledge of the frozen LLM $\mathcal{M}$ intact. For clarity, we summarize the mathematical notations in Table~\ref{tab:math}.

\begin{figure*}[!tp]
    \centering
    \setlength{\abovecaptionskip}{0cm}
    \setlength{\belowcaptionskip}{0cm}
\includegraphics[width=0.95\linewidth,height=0.45\linewidth]{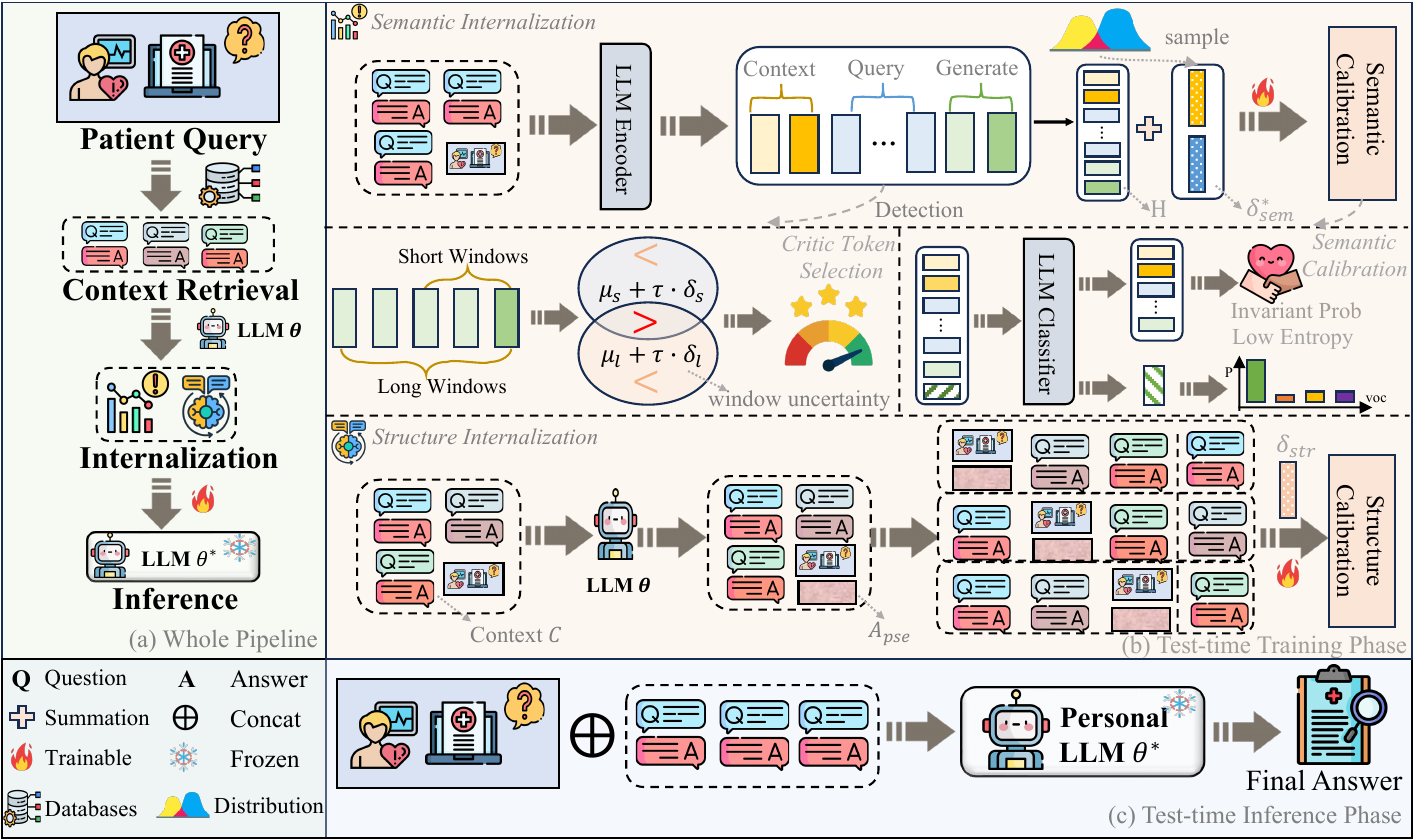}
    \caption{Overview of our DSC framework. The overall architecture of DSC is presented in (a), which outlines the comprehensive test-time training pipeline, while (b) provides a granular decomposition of our core dual-stream mechanism. DSC leverages two parallel streams to simultaneously refine the input from both semantic and structural dimensions. The Semantic Calibration Stream implements dynamic entropy detection and elimination. This process first localizes high-uncertainty tokens via long-short window entropy analysis and subsequently optimizes the model using dual objectives: an entropy loss to resolve ambiguity and a high-fidelity loss to preserve critical information, resulting in a semantically certain input representation. The Structure Calibration Stream runs in parallel to reconstruct the vital inferential links bridging the context and the clinical query. It employs an iterative, alternating framework where specialized support sets and meta-queries are dynamically synthesized for meta-optimization. Finally, (c) illustrates the test-time inference phase, where the model executes the final predictions using the parameters optimized specifically for the given query. $\theta^{*}$ denotes the LLM parameters integrated with  $\delta$.}
    \label{fig:frame}
\end{figure*}

\noindent\textbf{Solution Overview:}
Our solution proposes the DSC framework to achieve deep contextual internalization via fine-grained, output-level calibration at inference time, fundamentally advancing beyond passive knowledge exposure. DSC employs two distinct and parallel streams to simultaneously refine the input from semantic and structural perspectives.
The Semantic Calibration Stream refines conceptual integrity by addressing noise and ambiguity. This stream utilizes the dynamic entropy detection and elimination, which first identifies high-uncertainty tokens via long-short windows entropy analysis, and then optimizes the model using dual objectives—an entropy loss to reduce ambiguity and a recalibration factor loss to preserve critical information.
In parallel, the Structure Calibration Stream reconstructs the crucial inferential links between the context and the clinical query. This stream employs an iterative meta-learning framework that dynamically constructs specialized support sets and meta-queries to reconstruct the latent inferential trajectory. By forcing the model to actively navigate and map the dependencies between fragmented evidence, our approach transforms the context from a flat sequence into a structured logical backbone. This ensures that the final output is not merely a textual response, but a systematic derivation rooted in the structural necessity of the clinical evidence.
The unified, parallel DSC streams facilitate low-latency, inference-time adaptation, with the complete methodological architecture illustrated in Fig.~\ref{fig:frame}.

\subsection{Method}\label{sec:med}

In this subsection, we detail the DSC framework, engineered to achieve deep contextual internalization for clinical reasoning via fine-grained, output-level adaptation at inference time.

\subsubsection{Input Formulation and Context Retrieval}\label{sec:med:input}
The process begins with the construction of a contextual input $\mathbf{X}$. To bridge the gap between raw clinical inquiries and formal medical knowledge, and inspired by recent advances in query-centric adaptation~\cite{bpse}, we implement a two-stage query reformulation strategy: generating reasoning-oriented pseudo-labels followed by Top-$K$ context retrieval.

\noindent\textbf{Query Reformulation.}
Rather than relying on direct reasoning, we leverage the frozen LLM $\mathcal{M}$ to generate a reasoning-guided pseudo-label $\mathcal{A}_\text{pse}$. This label acts as an intermediate knowledge anchor that distills the latent intent of the raw clinical query $\mathcal{Q}$. Given an instruction template $\mathcal{I}_\text{pse}$, the model first synthesizes this pseudo-label, from which a refined, reformulated query $\mathcal{Q}'$ is subsequently extracted:
\begin{equation}\label{eq:5}
\mathcal{A}_\text{pse} = \mathcal{M}(\mathcal{I}_\text{pse}, \mathcal{Q}),
\end{equation}
\begin{equation}\label{eq:6}
\mathcal{Q}' = \text{Extract}(\mathcal{I}_\text{ref},\mathcal{Q}, \mathcal{A}_\text{pse}), 
\end{equation}
where $\mathcal{I}_\text{ref}$ denotes the reformulate instruction.
This ``generate-then-extract" mechanism effectively unpacks the complex clinical requirements of $\mathcal{Q}$, transforming it into a precise, reformulated probe $\mathcal{Q}'$ that is intrinsically optimized for high-precision retrieval.

\noindent\textbf{Context Retrieval.}
We derive the latent representation $\mathbf{H}_{\mathcal{Q}'}$ from the reformulated query $\mathcal{Q}'$ using the frozen encoder $\mathcal{E}$, i.e., $\mathbf{H}_{\mathcal{Q}'} = \mathcal{E}(\mathcal{Q'})$. Utilizing $\mathbf{H}_{\mathcal{Q}'}$, the context $\mathcal{C}$ is acquired via a {Top-$K$ retrieval} from the comprehensive training dataset $\mathcal{D}_{\text{train}}$:
\begin{equation}\label{eq:7}
\mathcal{C} = \text{Top-$K$}(\mathbf{H}_{\mathcal{Q}'}, \mathcal{D}_{\text{train}}) ,
\end{equation}
where $\mathcal{E}$ denotes the MedCPT \cite{emedcpt} used to compute cosine similarity for Top-$K$ evidence retrieval; a comprehensive analysis of alternative encoder architectures is provided in Section~\ref{sec:rob:plug}. The final input sequence is $\mathbf{X} = [\mathcal{C}; \mathcal{Q}']$. As mentioned in Eq.~\ref{eq:2}, we generate a final
hidden representation $\mathbf{H}\in\mathbb{R}^{d}$ of $\mathbf{X}$ at the penultimate layer. For the sake of notational simplicity, we deliberately eschew the use of $\mathcal{Q}'$ and continue to employ $\mathcal{Q}$ to represent the query, despite its enhanced informational density.

\subsubsection{Semantic Calibration Stream}\label{sec:med:sem}
This stream facilitates a deep comprehension and selective utilization of the evidence $\mathbf{X}$, effectively distilling core clinical insights while filtering the inherent semantic noise and ambiguity pervasive in the input. Rather than performing a discrete deletion of tokens, which may inadvertently disrupt linguistic coherence, this stream optimizes two latent correction vectors $\boldsymbol{\delta}_\text{sem}^{\mathcal{C}}$ and $\boldsymbol{\delta}_\text{sem}^{\mathcal{Q}}$. Please note that we apply distinct calibration vectors to context and query tokens. This decoupling enables independent control over their generative influence, filtering semantic noise without compromising the prompt's structural integrity. By perturbing the initial embedding $\mathbf{H}$ into a calibrated state $\mathbf{H}_\text{sem}^{*} = [\mathbf{H}_\mathcal{C} + \boldsymbol{\delta}_\text{sem}^\mathcal{C} \,;\, \mathbf{H}_\mathcal{Q} + \boldsymbol{\delta}_\text{sem}^\mathcal{Q}]$, the stream performs a fine-grained feature purification. This ensures the model selectively attends to the most salient evidence while suppressing the high-uncertainty generation. Ultimately, this mechanism compels the frozen LLM to perform a deep semantic internalization, isolating the high-value components within the context to anchor the final reasoning process.

\noindent\textbf{Critic Token Selection.}
To ensure the impact of these additional parameters is precisely guided, we implement an innovative dynamic entropy detection and elimination strategy. This approach is designed to isolate high-uncertainty tokens during the generation process, providing a key basis for the latent intervention. While point-wise entropy signals immediate uncertainty, it is prone to high-frequency noise and lacks contextual depth~\cite{etan2025uncertainty}. Furthermore, static thresholds fail to accommodate the high variability and informational richness of clinical narratives, as demonstrated in Table~\ref{tab:aba}.  To address this, we establish an adaptive thresholding mechanism based on two concurrent historical metrics: short-window entropy ($\mu_{\text{s}}$) and long-context entropy ($\mu_{\text{l}}$). 
A generated token $y_t$ is designated as a high-uncertainty point ($y_t \in \mathcal{U}$) only if its instantaneous entropy $H_t$ statistically deviates from both localized and global trends:
\begin{equation}\label{eq:8}
y_t \in \mathcal{U} \iff H_t > \mu_{\text{s}} + \tau  \cdot \sigma_{\text{s}} \quad \text{and} \quad H_t > \mu_{\text{l}} + \tau \cdot \sigma_{\text{l}} ,
\end{equation}
where $H_t = - \sum p(y_t) \log p(y_t)$ represents the Shannon entropy~\cite{slot} of the token distribution, and $\tau$ is sensitivity hyperparameters.  $\mu_{\text{s}}$ and its corresponding variance $\sigma_{\text{s}}$ are computed over a fixed-size window of the $N_{\text{short}}$ most recent tokens, capturing localized fluctuations in model confidence. In contrast, $\mu_{\text{l}}$ and $\sigma_{\text{l}}$ are cumulatively calculated across the entire (longer) generated sequence from the initial token to the current step $t$, reflecting the global stability of the reasoning trace. This dual-constraint mechanism ensures that the semantic calibration is highly selective; by synergizing localized sensitivity with global stability, we filter out only the most ambiguous representations, thereby minimizing global perturbation while maximizing the neutralization of semantic distractors. A comparative analysis of various uncertainty metrics is provided in Section~\ref{sec:rob:plug}.

\noindent\textbf{Calibration Optimization.}
The optimization seeks the optimal $\boldsymbol{\delta}_\text{sem}^{*}$ that minimizes uncertainty while preserving factual integrity. Naive entropy minimization in TLM~\cite{TTL} and SLOT~\cite{slot} risks catastrophic distributional collapse~\cite{entropy1}. The optimization must delicately balance ambiguity reduction with the preservation of critical non-uncertain information, for instance, by constraining the auto-regressive probability distributions of the context and query. We achieve this using a hybrid loss $\mathcal{L}_{\text{sem}}$:
\begin{equation}\label{eq:9}
\mathcal{L}_{\text{sem}} = \mathcal{L}_{\text{rcf}}(y^{<t}; \boldsymbol{\delta}_\text{sem}^{\mathcal{C}}, \boldsymbol{\delta}_\text{sem}^{\mathcal{Q}}) + \mathcal{L}_{\text{ent}}(y^{<t}; \boldsymbol{\delta}_\text{sem}^{\mathcal{C}}, \boldsymbol{\delta}_\text{sem}^{\mathcal{Q}}),
\end{equation}
where \textbf{$\mathcal{L}_{\text{ent}}$} directly targets the high uncertainty of $\mathcal{U}$ by minimizing the entropy of the perturbed predictive distribution:
    \begin{equation}\label{eq:10}       
    \mathcal{L}_{\text{ent}} = \frac{1}{|\mathcal{U}|} \sum_{y_t \in \mathcal{U}} H(P(\cdot|y^{<t}, \boldsymbol{\delta}_\text{sem}^{\mathcal{C}}, \boldsymbol{\delta}_\text{sem}^{\mathcal{Q}})).
    \end{equation}
To ensure that the adaptation does not degrade the model's foundational knowledge, the representation consistency factor $\mathcal{L}_{\text{rcf}}$ enforces distributional consistency for the tokens identified as certain ($\mathbf{X} \setminus \mathcal{U}$). Rather than a simple embedding constraint, $\mathcal{L}_{\text{rcf}}$ minimizes the divergence between the original and the adapted logit distributions, thereby preserving the semantics of the non-uncertain context:
\begin{equation}\label{eq:11} 
\mathcal{L}_{\text{rcf}} = \sum_{y_t \in \mathbf{X} \setminus \mathcal{U}} \text{D}_{\text{KL}} \left( P(y_t | \mathbf{H}_{y_t}) \parallel P(y_t | \mathbf{H}_{y_t} + \boldsymbol{\delta}_\text{sem}^{\mathcal{C}}+\boldsymbol{\delta}_\text{sem}^{\mathcal{Q}}) \right) ,
\end{equation}
where $P(y | \cdot)$ denotes the probability distribution generated by the frozen classification head $W_\text{logits}$. By penalizing shifts in the predicted probabilities for certain tokens, this objective ensures that the latent perturbation remains a targeted intervention. After optimization, the refined embedding $\mathbf{H}_\text{sem}^*$ constitutes the semantically certain representation $\mathcal{C}$. This formulation ensures that the output-level adapter selectively rectifies ambiguity without compromising the reasoning fidelity of the established clinical evidence.

\subsection{Structure Calibration Stream}\label{sec:med:str}
This stream runs in parallel to address the deficiency in the structural integration of scattered evidence. Rather than merely refining features, it enforces explicit structural alignment between the context $\mathcal{C}$ and the query $\mathcal{Q}$ by optimizing the structure calibration vector $\boldsymbol{\delta}_\text{str}$, thereby re-anchoring the model’s attention towards a logically coherent inferential path.

\noindent\textbf{Context Reformulation via Meta-Training.}
We achieve structural alignment by utilizing $\mathcal{C}$ as a support set $\mathcal{S}$ consisting of $K$ distinct samples, denoted as $\mathcal{S} = \{(\mathrm{x}_i, \mathrm{y}_i)\}_{i=1}^K$. To overcome the train-test mismatch prevalent in ICL, where the target query $\mathcal{Q}$ is structurally isolated, we incorporate $\mathcal{Q}$ directly into the structural learning process. We construct a {contextual alignment instance} $\mathcal{A}_\text{pse}$ by Eq.~\ref{eq:5} and augment the set: $\mathcal{S}' = \mathcal{S} \cup \{(\mathcal{Q}, \mathcal{A}_\text{pse})\}$.
Using a {leave-one-out strategy}~\cite{hmetac,MetaICL} over $\mathcal{S}'$, each pair $(\mathrm{x}_i, \mathrm{y}_i) \in \mathcal{S}'$ serves as a prediction target, while the remaining elements form the context. Through dynamic context permutation, the meta-training phase forces the LLM to capture invariant structural dependencies, facilitating robust knowledge transfer. The meta-training prompt is constructed as:
\begin{equation}\label{eq:12}
\mathcal{P}^{(i,l)}_{\text{meta}} = [\mathcal{I}_{\text{str}}; \text{Context}^{(l)}_i; \mathrm{x}_i] , 
\end{equation}
where $\text{Context}^{(l)}_i$ denotes the $l$-th structural permutation of the retrieved context for inquiry $\mathrm{x}_i$. Specifically, each version represents a distinct logical rearrangement of the scattered medical evidence, forcing the model to transcend fixed input orders. The complete meta-training dataset is $\mathcal{D}_{\text{meta}}^{i} = \left\{ (\mathcal{P}^{(i,l)}_{\text{meta}}, \mathrm{y}_i) | (\mathrm{x}_i, \mathrm{y}_i) \in S', l = 1, \dots, M \right\}$. To further boost the robustness of the learned structural path and enable the model to handle diverse reasoning directions (e.g., predicting cause from effect, or vice versa), we introduce the instance inversion augmentation.
For every original instance $(\mathrm{x}_i, \mathrm{y}_i) \in \mathcal{S}'$, we construct an inverted instance $(\mathrm{y}_i, \mathrm{x}_i)$ where the original label $\mathrm{y}_i$ becomes the new input (query), and the original input $\mathrm{x}_i$ becomes the new target (answer). Formally,
\begin{equation}\label{eq:13}
\mathcal{S}_{\text{inv}} = \{ (\mathrm{y}_i, \mathrm{x}_i) | (\mathrm{x}_i, \mathrm{y}_i) \in \mathcal{S}' \}.
\end{equation}
By training on the augmented dataset $\mathcal{S}' \cup \mathcal{S}_{\text{inv}}$, the model is compelled to learn bidirectional structural mappings. This ensures that the structure-aware query embedding $\mathbf{H}_\text{str}^*$ encodes dependencies that are robust to changes in the semantic role of the tokens. This is vital in clinical reasoning, where the structure of ``Symptoms $\to$ Diagnosis" must be related to ``Diagnosis $\to$ Symptoms/Tests". The overall meta-training dataset is expanded to $\mathcal{D}_{\text{meta}}^{i,*} = \mathcal{D}_{\text{meta}}^{i} \cup \mathcal{D}_{\text{meta}}^{i,\text{inv}}$.

\noindent\textbf{Calibration Optimization.}
The optimization of $\boldsymbol{\delta}_\text{str}$ minimizes the meta-training loss $\mathcal{L}_{\text{str}}$ over the augmented dataset $\mathcal{D}_{\text{meta}}^{i,*}$, compelling the query embedding to adaptively guide the structural path:
\begin{equation}\label{eq:14}
\mathcal{L}_{\text{str}}= - \sum_{(\mathcal{P},\mathrm{y}) \in \mathcal{D}_{\text{meta}}^{i,*}} \sum_{t=1}^{|\mathcal{D}_{\text{meta}}^{i,*}|}\log P(y_t | \mathcal{P}, y^{<t}, \boldsymbol{\delta}_\text{str}), 
\end{equation}
By directing the structural loss onto the input calibration vector $\boldsymbol{\delta}_\text{str}$, we effectively encode the required structural alignment into the query's latent space. This achieves the desired structural adaptation while preserving the frozen state of the base model's parameters. The final structure-aware query representation is then formulated as:
$\mathbf{H}_\text{str}^* = \mathbf{H} + \boldsymbol{\delta}_\text{str}$,
where $\boldsymbol{\delta}_\text{str}$ represents the optimized structural guidance that anchors the query to the retrieved clinical evidence. Our approach naturally conforms to meta-learning~\cite{hmeta-survey} principles: perturbed contexts serve as multiple learning pairs for the agent, with final optimization conducted via meta-query results.

\subsection{Test-time Training \& Inference}
This subsection delineates the complete test-time training and inference procedure. Each sample undergoes a sequential pipeline: first undergoing test-time training, followed by inference. Additionally, a comparative analysis of offline optimization variants is presented in Section~\ref{sec:rob:opt}.

\noindent\textbf{Training Objective.}
The final objective function $\mathcal{L}_{\text{dsc}}$ during the test-time training phase is a composite loss combining the two distinct calibration losses:
\begin{equation}\label{eq:15}
\mathcal{L}_{\text{dsc}} = \mathcal{L}_{\text{sem}}(\boldsymbol{\delta}_\text{sem}^{\mathcal{C}},\boldsymbol{\delta}_\text{sem}^{\mathcal{Q}}) + \gamma  \mathcal{L}_{\text{str}}(\boldsymbol{\delta}_\text{str}) ,
\end{equation}
where $\gamma$ is the trade-off weight. This composite objective realizes the core principle of dual-stream internalization. By simultaneously optimizing $\boldsymbol{\delta}_\text{sem}^{*}$ and $\boldsymbol{\delta}_\text{str}$ across their respective loss landscapes, the framework ensures both the clarity (semantic) and organization (structural) of the contextual evidence are maximized before the final inference. The main LLM weights $\theta$ remain frozen.

\noindent\textbf{Inference.}
During inference, the test-time training of $\boldsymbol{\delta}_\text{sem}^{*}$ and $\boldsymbol{\delta}_\text{str}$ is performed for a limited number of optimization steps $T_{\text{inf}}$ (e.g., $T_{\text{inf}} \ll 5$)~\cite{TTL,slot}. This rapid adaptation tailors the input representation to the specific test instance. The final, adaptively calibrated input $\mathbf{H}^{*}$ is constructed using the optimized $\boldsymbol{\delta}_\text{sem}^*$ and $\boldsymbol{\delta}_\text{str}$. The frozen LLM $\mathcal{M}$ then generates the final prediction:
\begin{equation}\label{eq:16}
\mathcal{A}_{\text{pred}} = \text{argmax}_{\mathcal{A}} P(\mathcal{A} | \mathbf{H}^*).
\end{equation}
This procedure ensures that the LLM performs the final reasoning step using a context that is both semantically certain and structurally optimized, leading to robust and accurate clinical reasoning.
The execution logic of our DSC framework is detailed in Algorithm~\ref{alg:dsc}.

\begin{algorithm}\small 
\caption{DSC for In-context Clinical Reasoning} 
\label{alg:dsc} 
\begin{algorithmic}[1] 
\REQUIRE Clinical Query $\mathcal{Q}$, Context $\mathcal{C}$, Frozen LLM $\mathcal{M}(\theta)$, Max Adaptation Steps $T_{\text{inf}}$;
\ENSURE Predicted Answer $\mathcal{A}_{\text{pred}}$;
\STATE Query reformulation $\mathcal{Q}'$ (Eq.~\ref{eq:6});
\STATE Retrieve context $\mathcal{C}$ via Top-$K$ retrieval (Eq.~\ref{eq:7});
\STATE Initialize calibration vectors $\boldsymbol{\delta}_\text{sem}^{\mathcal{C}}, \boldsymbol{\delta}_\text{sem}^{\mathcal{Q}}, \boldsymbol{\delta}_\text{str}$;
\STATE Initialize iteration $t=0$;
\WHILE {$t \leq T_{\text{inf}}$}

    \item[\textbf{ }] \textit{\# Semantic Calibration Stream}
    \STATE Identify high-uncertainty context tokens $\mathcal{U}$ using dual-window detectors (Eq.~\ref{eq:8});
    \STATE Compute semantic loss $\mathcal{L}_{\text{sem}}(\boldsymbol{\delta}_\text{sem}^{\mathcal{C}}, \boldsymbol{\delta}_\text{sem}^{\mathcal{Q}})$ (Eq.~\ref{eq:9});

    \item[\textbf{ }] \textit{\# Structure Calibration Stream}
    \STATE Construct meta-training dataset $\mathcal{D}_{\text{meta}}^{i}$ using leave-one-out and permutations (Eq.~\ref{eq:13});
    \STATE Compute structural loss $\mathcal{L}_{\text{str}}(\boldsymbol{\delta}_\text{str})$ (Eq.~\ref{eq:14});
    
    \item[\textbf{ }] \textit{\#  Unified Test-Time Training}
    \STATE Compute total loss $\mathcal{L}_{\text{dsc}}$(Eq.~\ref{eq:15});
    \STATE Update $\boldsymbol{\delta}_\text{sem}^{\mathcal{C}}, \boldsymbol{\delta}_\text{sem}^{\mathcal{Q}}, \boldsymbol{\delta}_\text{str}$ via $\nabla \mathcal{L}_{\text{dsc}}$;
    \STATE $t=t+1$;
\ENDWHILE

\item[\textbf{ }] \textit{\# Test-time Inference}
\STATE Construct final calibrated input $\mathbf{H}^* = \mathbf{H} + \boldsymbol{\delta}_\text{sem}^{\mathcal{C}} +\boldsymbol{\delta}_\text{sem}^{\mathcal{Q}}+\boldsymbol{\delta}_\text{str}$;
\STATE Generate prediction $\mathcal{A}_{\text{pred}} = \text{argmax}_{\mathcal{A}} P(\mathcal{A} | \mathbf{H}^*)$ using frozen $\mathcal{M}(\theta)$;

\RETURN $\mathcal{A}_{\text{pred}}$
\end{algorithmic}
\end{algorithm}

\section{Experiments}\label{sec:exp}
We conduct extensive experiments across three primary generative healthcare prediction tasks.

\noindent\textbf{Datasets \& Baselines.}
Our comprehensive evaluation spans three primary generative clinical reasoning tasks: Examination QA, Lay Summarization, and Clinical Diagnosis. Specifically, our evaluation covers seven datasets for Examination QA (including MedQA~\cite{datamedqa}, PubMedQA~\cite{datapubmed}, MedMCQA~\cite{datamcqa}, MedBullets~\cite{datamedbullets}, MMLU~\cite{datammlu}, MMLU-Pro~\cite{datammlupro}, and MedExQA~\cite{data-medexqa}.), three datasets for Lay Summarization (eLife~\cite{dataeLifePLOS}, Cochrane~\cite{datacochr}, and PLOS~\cite{dataeLifePLOS}), and three for Clinical Diagnosis (DiagnosisArena~\cite{datadiag}, ReDisQA~\cite{dataredi}, and MediQ~\cite{datamediq}). Please note that our experimental configurations vary by task: for Examination QA, we strictly adhere to the hard filtering settings established in~\cite{ztang2025,tags}; for Lay Summarization, we follow the same processing paradigm defined by~\cite{zyinhao2025}; and for Clinical Diagnosis, models and datasets are directly sourced from the Hugging Face Hub~\footnote{https://huggingface.co/} without additional task-specific modifications.

We categorize our comparative baselines into four distinct optimization paradigms: (i) Pure / Medical LLMs, including Qwen3-7B~\cite{qwen3}, DeepSeek-R1-7B~\cite{deepseekr1}, and Lingshu-7B~\cite{lingshu}; (ii) Training-dependent paradigms, such as SFT~\cite{sft} and GRPO~\cite{grpo}, requiring extensive pre-fitting on dedicated training sets; (iii) Test-time tuning-free paradigms, including ICL~\cite{ICL}, CoT~\cite{cot}, i-MedRAG~\cite{zXiong2025Improving}, and Ensemble (Major Voting)~\cite{majorvote}, which operate in long-context / few-shot modes without parameter updates; and (iv) Test-time learning paradigms, such as TLM~\cite{TTL}, TTT~\cite{TTT2025}, and SLOT~\cite{slot}, which utilize instance-specific adaptation similar to our DSC framework. Furthermore, we incorporate task-specific SOTA methods: MDAgents~\cite{MDAgents}, ColaCare~\cite{ColaCare, ColaCare2}, and TAGS~\cite{tags} for Examination QA; AgentSimp~\cite{AgentSimp} for Lay Summarization; and ColaCare~\cite{ColaCare, ColaCare2} and DiagRL~\cite{DiaRL} for Clinical Diagnosis. To ensure a fair comparison, all methods in paradigms (ii)–(iv) utilize the same LLM backbone as our framework, unless explicitly stated otherwise.
\begin{table}\small
\centering
\caption{Performance comparison: Examination QA (HARD VERSION~\cite{tags}). The pipeline integrates Qwen2.5-7B~\cite{qwen25} (LLM backbone), with further variations explored in Section~\ref{sec:rob:plug}.}
\label{tab:exp:qa}
\resizebox{0.5\textwidth}{!}{
\begin{tabular}{c|ccccccc} 
\toprule
Methods & MedQA & PubMedQA & MedMCQA & MedBullets & MMLU & MMLU-Pro & MedExQA  \\ 
\hline
Qwen2.5-7B~\cite{qwen25}     & 0.160      &   0.160         &  0.240      &   0.045          &   0.127      &  0.260      &   0.090      \\
Deepseek-R1-7B~\cite{deepseekr1}      & 0.180      &   0.230         &  0.260      &   0.134        &  0.260       &  0.100      &   0.210         \\
Lingshu-7B~\cite{lingshu}     & 0.190      &   0.170         &  0.260      &   {0.157}         & 0.273        &  0.210      &   0.210          \\
SFT~\cite{sft}    & 0.210 & 0.180 & 0.080 & 0.184 & 0.260 & 0.180 & 0.160 \\
GRPO~\cite{grpo}    & 0.230 & 0.240 & 0.230 & 0.134 & 0.150 & 0.180 & 0.180  \\
ICL~\cite{ICL}     & 0.230      &   0.190         &  0.200      &   0.168        &  0.219       &  0.250      &   0.150      \\
CoT~\cite{cot}    & 0.190      &   0.210         &  0.190      &   0.168       &  0.287       &  0.240      &   0.160         \\
Ensemble~\cite{majorvote}    & 0.220      &   0.200         &  0.270      &   0.112         & 0.136        &  0.290      &   0.130        \\
i-MedRAG~\cite{zXiong2025Improving} & 0.220      &   0.240         &  0.300      &   0.089         & 0.151        &  0.260      &   0.160          \\
TLM~\cite{TTL}     & 0.170      &   0.210         &  0.260      &   0.078         & 0.192        &  0.300      &   0.150         \\
SLOT~\cite{slot}    & 0.160      &   0.210         &  0.210      &   0.089          & 0.136        &  0.280      &   0.110          \\ 
TTT~\cite{TTT2025}    & 0.220      &   0.200         &  0.250      &   0.079          & 0.260        &  0.220      &   0.150         \\ 
MDAgents~\cite{MDAgents}     & 0.160      &   0.120         &  0.270      &   0.089       &  0.137       &  0.050      &   0.080          \\ 
ColaCare~\cite{ColaCare}~\cite{ColaCare2}    & 0.150      &   0.130         &  0.260      &   0.056          & 0.109        &  0.220      &   0.130         \\ 
TAGS~\cite{tags}     & 0.280      &   {0.250}         &  0.240      &   0.146         & \textbf{0.356}        &  0.250      &   0.160     \\
\hline
DSC    & \textbf{0.290}      &   \textbf{0.300}         &  \textbf{0.360}      &   \textbf{0.202}       &  0.301     &  \textbf{0.320}      &   \textbf{0.240}        \\
\bottomrule
\end{tabular}
}
\end{table}

\begin{table}\small
\centering
\caption{Performance comparison: Lay Summarization. In this task, we follow~\cite{zyinhao2025}. }
\label{tab:exp:sum}
\resizebox{0.5\textwidth}{!}{
\begin{tabular}{c|ccc||ccc||ccc} 
\toprule
Methods & \multicolumn{3}{c||}{Cochrane}         & \multicolumn{3}{c||}{eLife}         & \multicolumn{3}{c}{PLOS}    \\ 
\cline{2-10}
Metrics & ROUGE-1    & ROUGE-L    & SARI    & ROUGE-1    & ROUGE-L    & SARI   & ROUGE-1    & ROUGE-L    & SARI  \\ 
\hline
Qwen2.5-7B~\cite{qwen25}      & 0.397 & 0.372 & 0.384 & 0.350 & 0.322 & 0.433 & 0.370 & 0.348 & 0.394  \\
Deepseek-R1-7B~\cite{deepseekr1}     & 0.400 & 0.369 & 0.383 & 0.368 & 0.349 & 0.437 & 0.410 & 0.372 & 0.367  \\
Lingshu-7B~\cite{lingshu}    & 0.385 & 0.357 & 0.383 & 0.312 & 0.291 & 0.428 & 0.357 & 0.329 & 0.384  \\
SFT~\cite{sft}    & 0.435 & 0.400 & 0.382 & 0.423 & 0.409 & 0.405 & 0.415 & 0.371 & 0.396  \\
GRPO~\cite{grpo}    & 0.410 & 0.380 & 0.389 & 0.405 & 0.382 & 0.443 & 0.409 & 0.387 & 0.397  \\
ICL~\cite{ICL}    & 0.431 & 0.403 & 0.396 & 0.431 & 0.406 & 0.449 & 0.398 & 0.372  & 0.401 \\
CoT~\cite{cot}    & 0.418 & 0.352 & 0.396 & 0.420 & 0.397 & 0.446 & 0.390 & 0.366 & 0.403  \\
Ensemble~\cite{majorvote}  & 0.367 & 0.304  & 0.350 & 0.290 & 0.257 & 0.397   & 0.280 & 0.267 & 0.358  \\
i-MedRAG~\cite{zXiong2025Improving} & 0.327      &   0.276         &  0.366      &   0.413         & 0.390        &  0.427      &   0.325       &  0.298      &   0.374     \\
TLM~\cite{TTL}    & 0.354 & 0.331 & 0.386 & 0.372 & 0.366 & 0.430 & 0.355 & 0.334 & 0.393  \\ 
SLOT~\cite{slot}    & 0.353 & 0.330 & 0.386 & 0.352 & 0.334 & 0.435 & 0.373 & 0.350 & 0.463  \\ 
TTT~\cite{TTT2025}    & 0.408 & 0.382 & 0.397 & 0.403 & 0.379 & 0.446 & 0.391 & 0.366  & 0.402 \\ 
AgentSimp~\cite{AgentSimp}    & 0.382 & 0.352 & 0.395 & 0.421 & 0.358 & 0.427 & 0.391 & 0.353 & 0.365  \\
\hline
DSC    & \textbf{0.453} & \textbf{0.423} & \textbf{0.402} & \textbf{0.448} & \textbf{0.430} & \textbf{0.453} & \textbf{0.442} & \textbf{0.416} & \textbf{0.444}  \\
\bottomrule
\end{tabular}
}
\end{table}

\begin{table}
\centering
\caption{Performance comparison: Clinical Diagnosis.}
\label{tab:exp:diag}
\resizebox{0.5\textwidth}{!}{
\begin{tabular}{c|cc||cc||cc} 
\toprule
Methods & \multicolumn{2}{c||}{MediQ}         & \multicolumn{2}{c||}{ReDisQA}         & \multicolumn{2}{c}{DiagnosisArena}    \\ 
\cline{2-7}
Metrics & ACC    & ROUGE-L      & ACC    & ROUGE-L     & ACC    & ROUGE-L  \\ 
\hline
Qwen2.5-7B~\cite{qwen25}      & 0.593 & 0.518  & 0.595 & 0.561 & 0.293 & 0.389   \\
Deepseek-R1-7B~\cite{deepseekr1}     & 0.358 & 0.450  & 0.389 & 0.257  & 0.250  & 0.163  \\
Lingshu-7B~\cite{lingshu}    & 0.589 & 0.467  & 0.566  & 0.529 & 0.423 & 0.164   \\
SFT~\cite{sft}    & 0.533  & 0.567 & 0.588  & 0.592 & 0.444  & 0.516  \\
GRPO~\cite{grpo}    & 0.585  & 0.577 & 0.544  & 0.597 & 0.444  & 0.521  \\
ICL~\cite{ICL}    & 0.527  & 0.619 & 0.573  & 0.581 & 0.380   & 0.479 \\
CoT~\cite{cot}    & 0.538 & 0.602  & 0.632  & 0.563 & 0.326 & 0.383   \\
Ensemble~\cite{majorvote}    & 0.608 & 0.603  & 0.647 & 0.314  & 0.369 & 0.381   \\
i-MedRAG~\cite{zXiong2025Improving} & 0.603               &  0.570      &   0.661         & 0.450             &   0.271            &   0.279     \\
TLM~\cite{TTL}    & 0.612 & 0.622 & 0.653 & 0.577 & 0.369 & 0.448  \\ 
SLOT~\cite{slot}    & 0.575 & 0.556 & 0.637 & 0.556 & 0.358  & 0.425  \\ 
TTT~\cite{TTT2025}    & 0.600 & 0.487 & 0.573 & 0.550 & 0.369   & 0.413 \\ 
ColaCare~\cite{ColaCare}\cite{ColaCare2}    & 0.607  & 0.580 & 0.588  & 0.479 & 0.402  & 0.344  \\
DiagRL~\cite{DiaRL}    & 0.545  & 0.619 & 0.566  & 0.587 & 0.347 & 0.422   \\
\hline
DSC    & \textbf{0.632} & \textbf{0.634} & \textbf{0.677} & \textbf{0.610} & \textbf{0.456} & \textbf{0.532}  \\
\bottomrule
\end{tabular}
}
\end{table}

\noindent\textbf{Implementation Details \& Metrics.}
We implement the DSC framework and all competitive baselines using PyTorch 2.0 and the Hugging Face Transformers library, conducting experiments on a hardware configuration featuring an Intel Xeon CPU and eight NVIDIA A800 GPUs. We select Qwen2.5-7B-Instruct~\cite{qwen25} as the frozen backbone LLM $\mathcal{M}(\theta)$, which remains fixed throughout the test-time adaptation phase. For retrieval (Eq.~\ref{eq:7}), we employ MedCPT~\cite{emedcpt} for Examination QA and E5~\cite{e5} for the other two tasks as our embedding models, utilizing FAISS~\cite{edouze2025faiss} to index the biomedical corpus. The core trainable components, the correction vectors $\boldsymbol{\delta}_\text{sem}^{*}$ and $\boldsymbol{\delta}_\text{str}$ (which match the LLM's hidden size), are initialized to zero vectors before the optimization of each test instance. Adaptation is performed for a minimal number of steps, ${T_{\text{inf}}=5}$. We employ the AdamW optimizer, setting the learning rates to ${1e-2}$. The final loss balancing hyperparameter (Eq.~\ref{eq:15}) are configured as ${\gamma=0.05}$, weighting the structural enforcement. For the Semantic Calibration Stream, the dynamic threshold parameters are set to ${\tau=3}$, with a short-window size of ${N_{\text{short}}=25}$. The Top-$K$ retrieval size is fixed at ${K=3}$ unless otherwise specified. Please note that for lay summarization, we only utilize 200 randomly sampled training instances for a warm start, a scale significantly smaller than SFT or GRPO (vs. thousands in these two). All these key parameters are determined based on the hyperparameter analysis detailed in Section~\ref{sec:hyper}. 
For Examination QA, following~\cite{MedAgents,tags}, we use Accuracy (ACC) to measure discrete reasoning precision.
For Lay Summarization, adhering to~\cite{zyinhao2025}, we employ ROUGE-1, ROUGE-L, and SARI to assess linguistic and structural fidelity.
For Clinical Diagnosis, following~\cite{datadiag,MedAgents,tags}, we utilize ACC alongside ROUGE-L to capture semantic alignment with expert ground truths.





\noindent\textbf{Overall Results.}  As demonstrated in Tables~\ref{tab:exp:qa}, \ref{tab:exp:sum}, and \ref{tab:exp:diag}, DSC consistently outperforms baseline models across diverse tasks and metrics, particularly showing significant gains in complex, multi-hop reasoning tasks where contextual fidelity is paramount. While inferior to the best baseline on MMLU, our model remains competitive. We attribute this to only 5 training samples being used as retrieval sources, which may cause inherent homogeneity.

Deepseek-R1-7B exhibits limited performance in Examination QA and Clinical Diagnosis, stemming from the knowledge gap and the inherent entropy preference of general models to exhibit high predictive uncertainty as discussed in Fig.~\ref{fig:motiv:ent}. Performance sees a moderate increase with ICL and GRPO, as the introduction of external context /adaptation mitigates the knowledge gap. However, both ICL and i-MedRAG are limited by their passive knowledge exposure: the LLM is forced to process raw context without an internal mechanism to filter noise or align the evidence structure. This leads to susceptibility to the ``loss-in-the-middle" problem, where crucial evidence is overlooked due to the model focusing its limited attention budget on noisy tokens. TLM and SLOT, while offering dynamic adaptation, often suffer from noise amplification, as their full-token optimization can inadvertently reinforce misalignments. In  contrast to passive methods, our DSC yields the most robust performance gains by orchestrating active, dual-stream knowledge internalization. The Semantic Calibration Stream reduces noise amplification, a common failure mode in context-driven generation, by isolating high-uncertainty tokens. Concurrently, the Structure Calibration Stream enforces a rigorous inferential bridge between in-context evidence and final predictions, effectively resolving structural ambiguity and facilitating knowledge transfer.

Across the hard Examination QA tasks, we observe that the MedMCQA dataset yields the highest overall performance. This peak is likely attributable to its streamlined contextual density and high alignment with the clinical textbook knowledge encoded within the pre-trained weights or guidelines of models like Lingshu and i-MedRAG.
In Lay Summarization, Cochrane demonstrates superior results compared to other datasets due to its structured multi-layered hierarchy. This standardized clinical reporting structure serves as a natural architectural bridge to the latent reasoning logic of high-capacity models like DeepSeek-R1, thereby maximizing generative coherence.
In Clinical Diagnosis, algorithms such as CoT and TLM universally perform better on ReDisQA compared to DiagnosisArena. This disparity is driven by ReDisQA’s constrained context space and its integration of critical medication metadata, which effectively simplifies the model's decision-making manifold. Simultaneously, the comprehensive patient profiles provided by the latter impose a significant computational and reasoning burden on the models. While these profiles offer high-fidelity clinical signals, their inherent complexity and high informational density necessitate advanced cross-referencing capabilities, which often exceed current model limits and lead to a noticeable degradation.

In terms of task complexity, Lay Summarization, which necessitates free-text and long-form synthesis, proves the most formidable challenge. Baseline models exhibit degradation in chronological coherence, frequently succumbing to catastrophic error propagation. In such high-entropy scenarios, standard RAG and Ensemble fail to navigate dense contextual dependencies. Clinical Diagnosis is an intermediate-complexity tier, particularly for TTL-based algorithms, as it requires precise probabilistic balancing across evolving dialogue states. In this setting, our Semantic Calibration Stream serves as a critical stabilization mechanism, ensuring the model remains resilient against noisy differential diagnoses or conflicting symptomatic reports. By ensuring both semantic certainty and structural synthetics, this dual-stream filtration empowers the frozen LLM to execute high-confidence inference, thereby catalyzing the substantial performance gains across all context-dependent clinical tasks.

\section{Model Analysis and Robust Testings}\label{sec:rob}
We conduct numerous robustness experiments to provide a more in-depth analysis.  Without loss of generality, we use MedQA, eLife, and DianosisArena for examination.

\subsection{Ablation Studies}\label{sec:rob:aba}
We conduct extensive ablation analyses on the designed submodules while keeping other components consistent to validate the effectiveness of each element within the DSC framework. As shown in Table~\ref{tab:aba}, a performance decline is observed with any ablated variant, demonstrating the indispensability of each submodule.
The {DSC-NC} variant isolates the impact of the retrieved context by removing $\mathcal{C}$ entirely. This setup degenerates the DSC framework into a localized test-time calibration restricted to the query embedding, nullifying the Structure Calibration Stream due to the absence of external evidence anchors. The resulting large performance gap highlights that DSC's efficacy is rooted not merely in query refinement, but in the synergistic purification and structural alignment of external knowledge. 
DSC-NSW relies solely on the long-context entropy average, making it overly conservative and slow to react to long/short uncertainty spikes; this results in a $6\%$ performance degradation on eLife. Conversely, DSC-NLW relies only on the short-window local entropy, leading to unstable and overly aggressive intervention, as it frequently misidentifies natural complexity as uncertainty, causing a $3.4\%$ drop on MedQA. These results validate our core insight that the dual-window, dynamic threshold approach is necessary for precise and stable noise detection.
{DSC-NR} removes the $\mathcal{L}_{\text{rcf}}$ loss from the Semantic Stream. $\mathcal{L}_{\text{rcf}}$ encourages stable learning by anchoring the current correction vector to its prior state. Ablating it leads to an unstable optimization trajectory for $\boldsymbol{\delta}_\text{sem}^{*}$ and a performance decrease, demonstrating the necessity of this regularization term to prevent over-calibration during the sparse, high-magnitude intervention.
{DSC-NS} and {DSC-NST}  directly ablate the two main streams of our framework. The {DSC-NS} variant removes the Semantic Calibration Stream, stripping the framework of its ability to resolve latent ambiguity in the query and context. This shift from active refinement to passive knowledge exposure precipitates a significant performance drop. Similarly, DSC-NST removes the Structure Calibration Stream, preventing the model from aligning the query to the required inferential structure, which significantly undermines its in-context robustness and leads to a $4.6\%$ performance degradation on eLife. 

In summary, these extensive ablation experiments robustly confirm that our core hypothesis that dual-stream input adaptation is critical for achieving state-of-the-art performance in complex contextual reasoning tasks. 
\begin{table}[!ht]
\centering
\setlength{\belowcaptionskip}{-0.1cm}   
\caption{Ablation Study. -NC ablates the retrieved context $\mathcal{C}$; -NSW/-NLW ablate the short/long entropy windows detection using fixed threshold, respectively; -NR ablates the $\mathcal{L}_{\text{rcf}}$ regularization term; -NS ablates the entire Semantic Calibration Stream; and -NST ablates the Structure Calibration Stream.} 
\label{tab:aba}
\resizebox{0.48\textwidth}{!}{
\begin{tabular}{c|c|cccccc|c} 
\hline
Algorithms                & Metric   & -NC & -NSW & -NLW & -NR & -NS & -NST & \textbf{DSC}  \\ 
\hline
\multirow{1}{*}{MedQA} & ACC    & 0.190  & 0.210   & 0.280   & 0.260   & 0.240  & 0.210  & \textbf{0.290}      \\
\hline
\multirow{3}{*}{eLife} & ROUGE-1   & 0.419  & 0.420   & 0.439   & 0.432 & 0.427  & 0.427    & \textbf{0.448}        \\
& ROUGE-L   & 0.392  & 0.408   & 0.412  & 0.403  & 0.401   & 0.393   & \textbf{0.430}        \\
& SARI   & 0.424  & 0.451   & 0.449   & 0.451   & 0.450 & 0.433    & \textbf{0.453}        \\
\hline
\multirow{2}{*}{DiagnosisArena} & ACC   & 0.391   & 0.424  & 0.391   & 0.444   & 0.424  & 0.413  & \textbf{0.456}       \\
& ROUGE-L   & 0.465  & 0.479   & 0.518   & 0.523 & 0.508    & 0.491  & \textbf{0.532}        \\
\hline
\end{tabular}}
\end{table}

\subsection{Plug-in Examination}\label{sec:rob:plug}
\begin{figure}[!h] 
\centering
\subfigure[MedQA (ACC)]{
\begin{minipage}[t]{0.325\linewidth}
\centering
\includegraphics[width=\linewidth,height=0.95\linewidth]{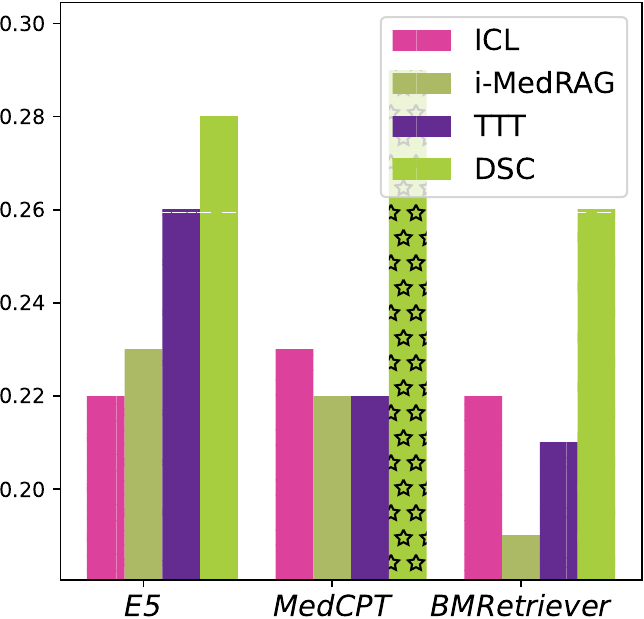}\label{fig:prm:medqa:acc}
\end{minipage}%
}%
\subfigure[eLife (R-L)]{
\begin{minipage}[t]{0.325\linewidth}
\centering
\includegraphics[width=\linewidth,height=0.95\linewidth]{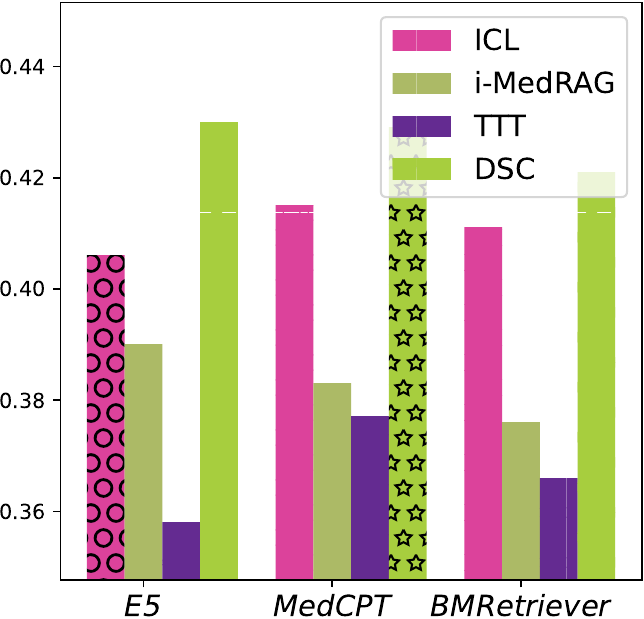}\label{fig:prm:elife:rl}
\end{minipage}%
}%
\subfigure[DiagnosisArena(ACC)]{
\begin{minipage}[t]{0.325\linewidth}
\centering
\includegraphics[width=\linewidth,height=0.95\linewidth]{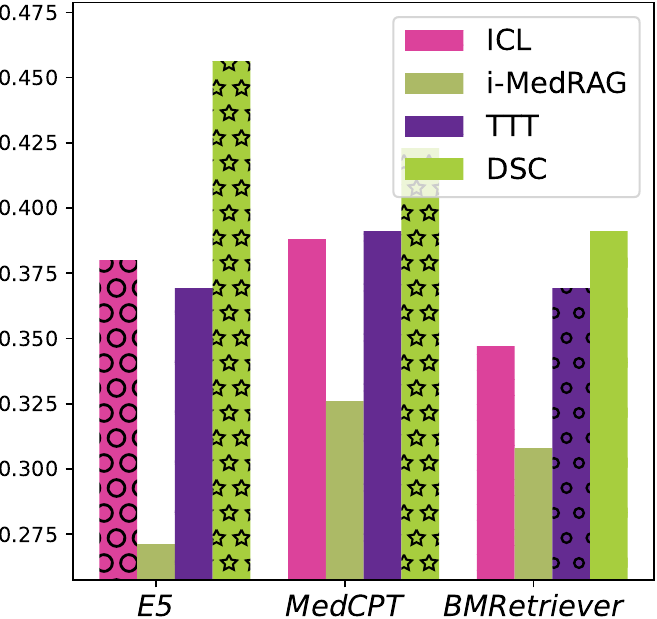}\label{fig:prm:diag:acc}
\end{minipage}%
}%
\centering
\setlength{\abovecaptionskip}{-0.15cm}   
\setlength{\belowcaptionskip}{-0.1cm}   
\caption{Comparison under diverse retrievers. We employ the popular BMRETRIEVER~\cite{embxu}, E5~\cite{e5}, and MedCPT~\cite{emedcpt}. We compare retrieval-dependent baselines, including ICL, i-MedRAG, and TTT. }
\label{fig:plug:retriever} 
\vspace{-0.3cm}
\end{figure}

\noindent\textbf{Context Retrievers.} 
The choice of retrieval model directly influences the initial quality of the context $\mathcal{C}$ (Eq.~\ref{eq:7}), thereby affecting the workload of the two calibration streams. We evaluate three different state-of-the-art embedding models: E5~\cite{e5}; {BMRetriever}~\cite{embxu}, a domain-specific model fine-tuned on clinical queries; and {MedCPT}~\cite{emedcpt,zXiong2025Improving}, a strong medical-purpose embedding. As depicted in Fig.~\ref{fig:plug:retriever}, all retrievers enable DSC to achieve strong performance, indicating the framework's robustness against varying context quality. Specifically, {MedCPT} does not always yield the largest performance gain, providing only marginal improvement over the {BMRetriever} embedding. This observation demonstrates a key advantage of the DSC framework: its Semantic Calibration Stream effectively mitigates the impact of suboptimal or noisy contexts, preventing the LLM's final prediction from being unduly influenced by irrelevant evidence, even when extracted by a general-purpose retriever. 

\begin{figure}[!h] 
\centering
\subfigure[MedQA (ACC)]{
\begin{minipage}[t]{0.325\linewidth}
\centering
\includegraphics[width=\linewidth,height=0.95\linewidth]{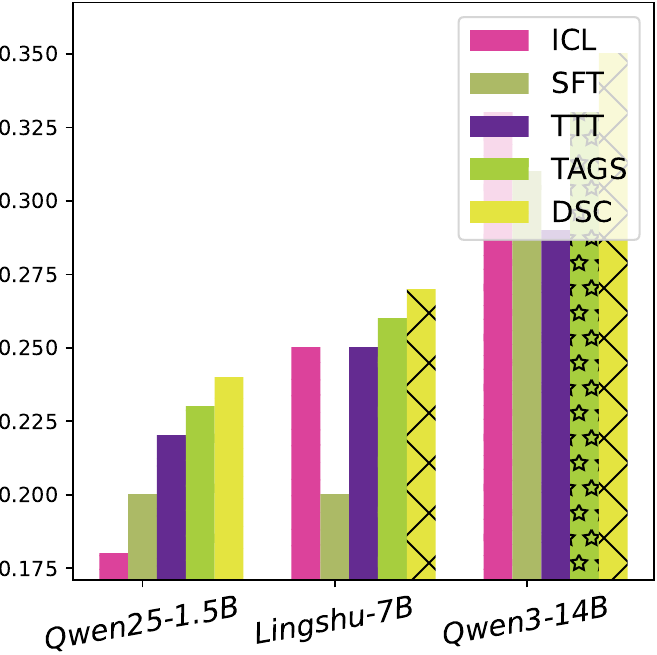}\label{fig:llm:medqa:acc}
\end{minipage}%
}%
\subfigure[eLife (R-L)]{
\begin{minipage}[t]{0.325\linewidth}
\centering
\includegraphics[width=\linewidth,height=0.95\linewidth]{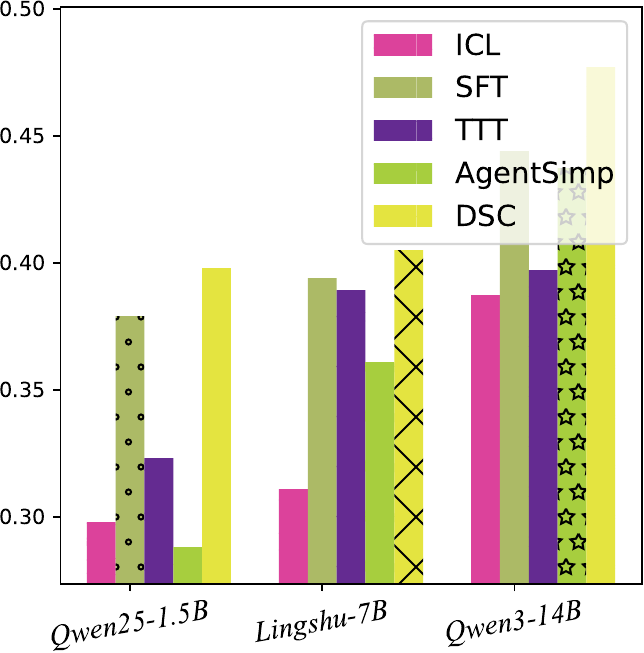}\label{fig:llm:elife:rl}
\end{minipage}%
}%
\subfigure[DiagnosisArena(ACC)]{
\begin{minipage}[t]{0.325\linewidth}
\centering
\includegraphics[width=\linewidth,height=0.95\linewidth]{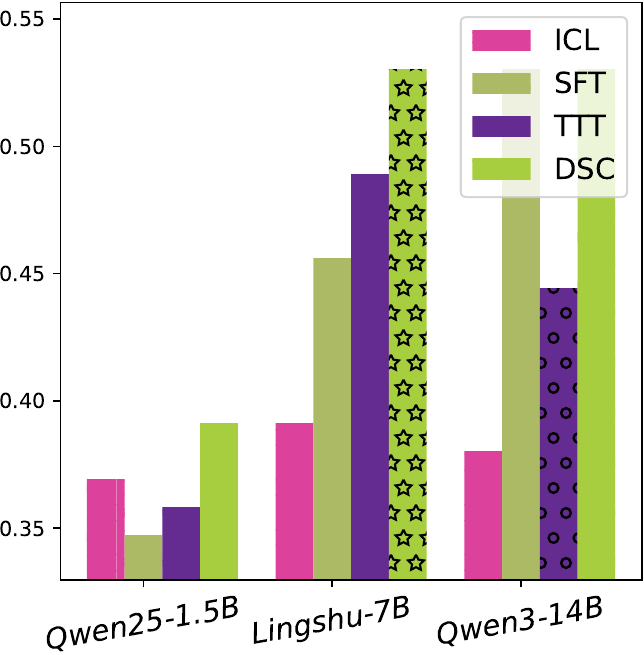}\label{fig:llm:diag:acc}
\end{minipage}%
}%
\centering
\setlength{\abovecaptionskip}{-0.15cm}   
\setlength{\belowcaptionskip}{-0.1cm}   
\caption{Comparison under diverse LLMs. We employ Qwen2.5-1.5B~\cite{qwen25}, Lingshu-7B~\cite{lingshu}, and Qwen3-14B~\cite{qwen3}. We select ICL, SFT, i-MedRAG, and TTT as representative baselines. We additionally incorporate TAGS for MedQA and AgentSimp for eLife due to their competitive performance.} 
\label{fig:plug:llm}
\end{figure}

\noindent\textbf{LLM Backbones.}
We test the portability and efficiency of DSC by varying the base LLM backbone, including smaller LLM ({Qwen2.5-1.5B}~\cite{qwen25}), Medical LLM (Lingshu-7B~\cite{lingshu}), and large reasoning LLMs ({Qwen3-14B}~\cite{qwen3}). The backbone determines the fundamental reasoning capacity and the quality of initial embeddings ($\mathbf{H}$). As shown in Fig.~\ref{fig:plug:llm}, increasing the model size generally correlates with performance improvements, with {Qwen3-14B} achieving the highest score. However, the improvement gap between the 7B model and the 14B model is notably small, and the {Qwen2.5-1.5B} model, when augmented with DSC, significantly outperforms its few-shot ICL and competes effectively with much larger baselines. This demonstrates that for complex RAG tasks, adaptive input calibration is a highly efficient alternative to scaling up the base model parameters, proving DSC's value for resource-constrained clinical environments. We also observe that utilizing Lingshu-7B variants yields only marginal improvements over the core DSC architecture in Section~\ref{sec:exp}. This minimal variance suggests that DSC's efficacy is largely decoupled from domain-specific pre-training. Instead, the framework functions as a robust test-time enhancer, prioritizing the dynamic internalization of query-context relationships over a reliance on static internal weights.

\begin{figure}[!h] 
\centering
\subfigure[MedQA (ACC)]{
\begin{minipage}[t]{0.325\linewidth}
\centering
\includegraphics[width=\linewidth,height=0.95\linewidth]{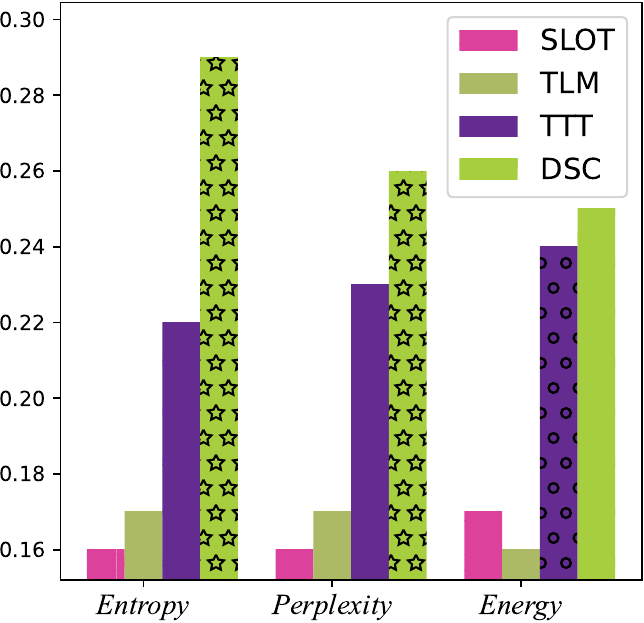}\label{fig:unc:medqa:acc}
\end{minipage}%
}%
\subfigure[eLife (R-L)]{
\begin{minipage}[t]{0.325\linewidth}
\centering
\includegraphics[width=\linewidth,height=0.95\linewidth]{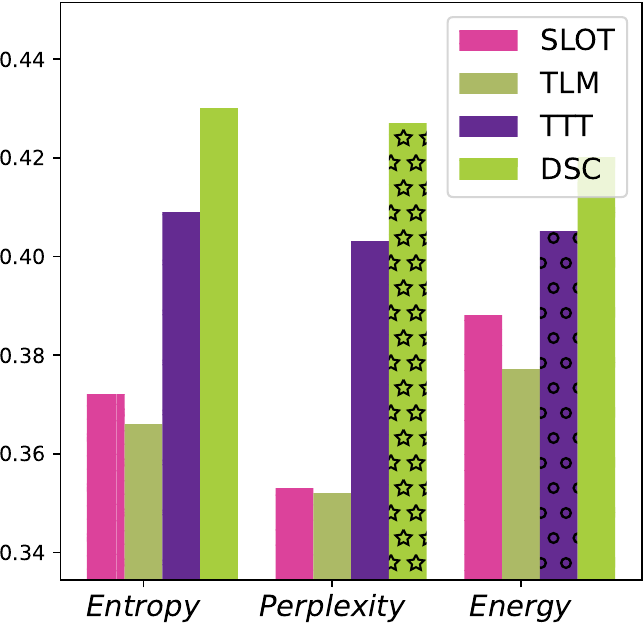}\label{fig:unc:elife:rl}
\end{minipage}%
}%
\subfigure[DiagnosisArena(ACC)]{
\begin{minipage}[t]{0.325\linewidth}
\centering
\includegraphics[width=\linewidth,height=0.95\linewidth]{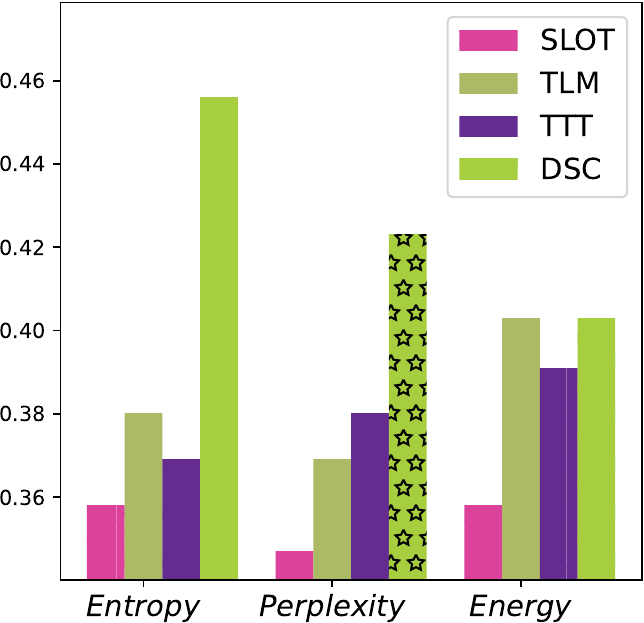}\label{fig:unc:diag:acc}
\end{minipage}%
}%
\centering
\setlength{\abovecaptionskip}{-0.15cm}   
\setlength{\belowcaptionskip}{-0.1cm}   
\caption{Comparison under diverse uncertainty estimation. Following~\cite {etan2025uncertainty}, we employ Entropy, Perplexity, and Energy. TTL-based baselines are included.}
\label{fig:plug:uncertainty}
\end{figure}

\noindent\textbf{Uncertainty Estimations.}
The dynamic entropy detection relies on accurate quantification of uncertainty. We test different metrics~\cite{eshorinwa2025survey} for high-uncertainty token identification $\mathcal{U}$: standard {Perplexity}, {Entropy} (our choice in Eq.~\ref{eq:10}), and Energy. The metric choice dictates which tokens are targeted by the stream optimization. As depicted in Fig.~\ref{fig:plug:uncertainty}, Entropy yields the competitive potential performance. Among practical, inference-time metrics, using {Entropy} significantly outperforms Perplexity. This is because Perplexity provides a general measure of sequence fluency, which is often too broad and fails to localize prediction ambiguity effectively. In contrast, Entropy measures the dispersion of the next-token probability distribution, directly corresponding to the model's predictive certainty at token $t$. This localization is essential for the stream to selectively apply $\boldsymbol{\delta}_\text{sem}^{*}$ and maximize the impact of the $\mathcal{L}_{\text{ent}}$ without corrupting stable context areas.

\subsection{Online vs. Offline Test-time Optimization}\label{sec:rob:opt}

\begin{figure}[!h] 
\centering
\subfigure[MedQA (ACC)]{
\begin{minipage}[t]{0.325\linewidth}
\centering
\includegraphics[width=\linewidth,height=0.95\linewidth]{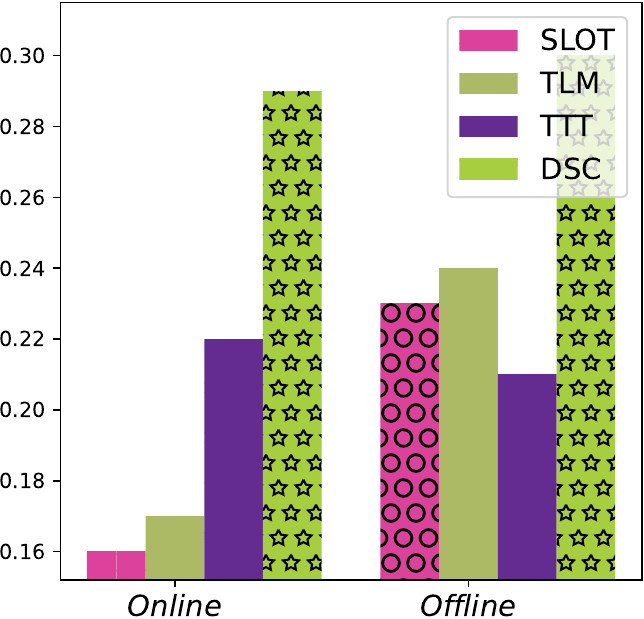}\label{fig:onf:medqa:acc}
\end{minipage}%
}%
\subfigure[eLife (R-L)]{
\begin{minipage}[t]{0.325\linewidth}
\centering
\includegraphics[width=\linewidth,height=0.95\linewidth]{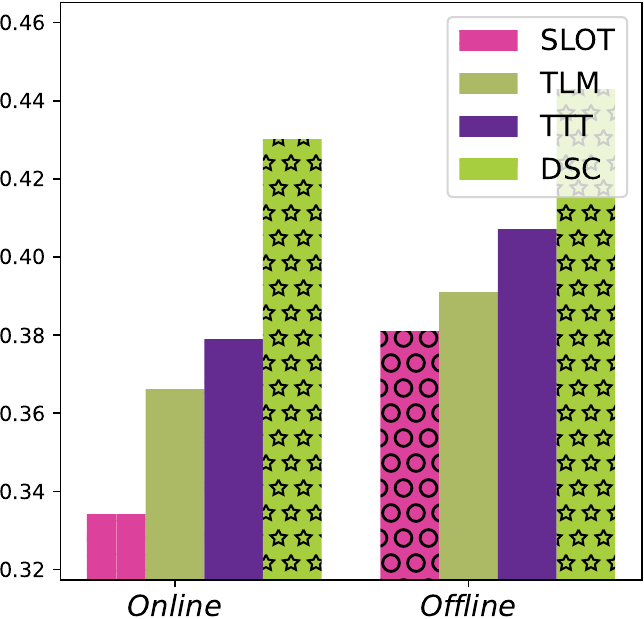}\label{fig:onf:elife:rl}
\end{minipage}%
}%
\subfigure[DiagnosisArena(ACC)]{
\begin{minipage}[t]{0.325\linewidth}
\centering
\includegraphics[width=\linewidth,height=0.95\linewidth]{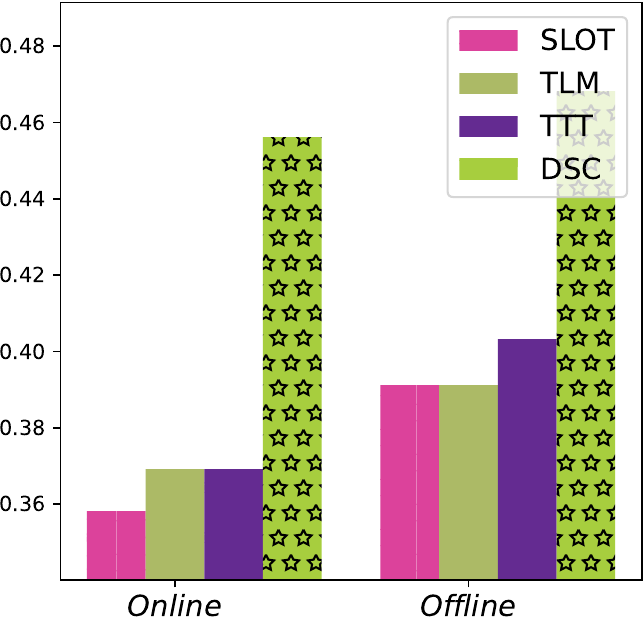}\label{fig:onf:diag:acc}
\end{minipage}%
}%
\centering
\setlength{\abovecaptionskip}{-0.15cm}   
\setlength{\belowcaptionskip}{-0.1cm}   
\caption{Online vs. Offline test-time optimization. Online methods tailor the model to each test query individually prior to inference, whereas offline methods conduct a one-time optimization on the entire test set before evaluating queries sequentially. TTL-based baselines are included.}
\label{fig:onf}
\end{figure}

We analyze the performance characteristics of DSC under two distinct test-time optimization paradigms, defined by the sequence of adaptation and evaluation~\cite{TTL,etan2025uncertainty}. In the {online optimization} scenario, adaptation is interleaved: for a given test sample $\mathbf{X}_i$, the correction vectors ($\boldsymbol{\delta}_\text{sem}^{*}, \boldsymbol{\delta}_\text{str}$) are optimized for $T_{\text{inf}}$ steps using $\mathcal{L}_{\text{dsc}}$ (Eq.~\ref{eq:15}), and predictions are immediately made on $\mathbf{X}_i$, discarding the adaptation before processing $\mathbf{X}_{i+1}$. Conversely, in the {offline optimization} approach, the model iterates through all test batches ($\mathbf{X}_{1}, \dots, \mathbf{X}_{N}$) for adaptation, and predictions are only made on the entire test set after all batches have been processed. As shown in Fig.~\ref{fig:onf}, the online optimization setting achieves performance that is highly competitive with the offline setting. This performance gap highlights the strong instance-specificity of the DSC framework. Because the mechanism is designed to adapt to the unique semantic noise and structural requirements of each input instance, the benefits derived from optimizing the full batch sequence (offline) do not significantly transfer or generalize across test instances. This also proves that the DSC  performs effective, rapid, and isolated adaptation, making it ideal for real-world online inference where low latency and batch independence are critical requirements.

\subsection{Out-of-Distribution Examination}\label{sec:rob:ood} 
\begin{figure}[!h] 
\centering
\subfigure[MedQA$\rightarrow$PubMedQA]{
\begin{minipage}[t]{0.49\linewidth}
\centering
\includegraphics[width=\linewidth,height=0.8\linewidth]{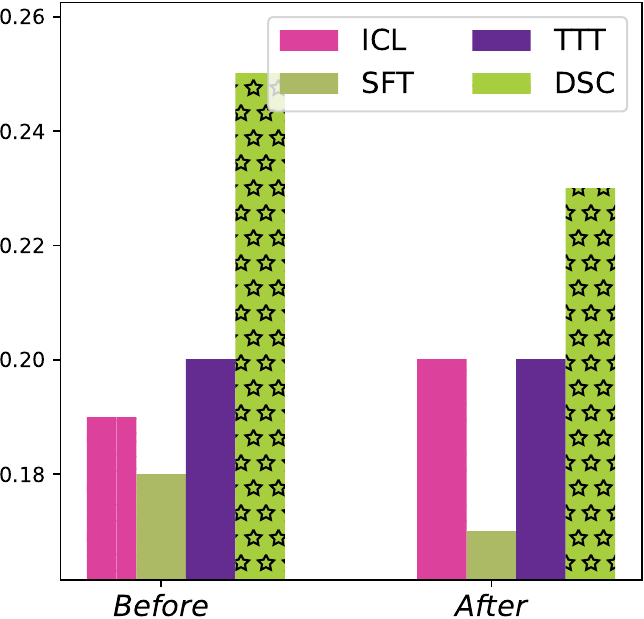}\label{fig:ood:data}
\end{minipage}%
}%
\subfigure[eLife$\rightarrow$DiagnosisArena-MCQ]{ 
\begin{minipage}[t]{0.49\linewidth}
\centering
\includegraphics[width=\linewidth,height=0.8\linewidth]{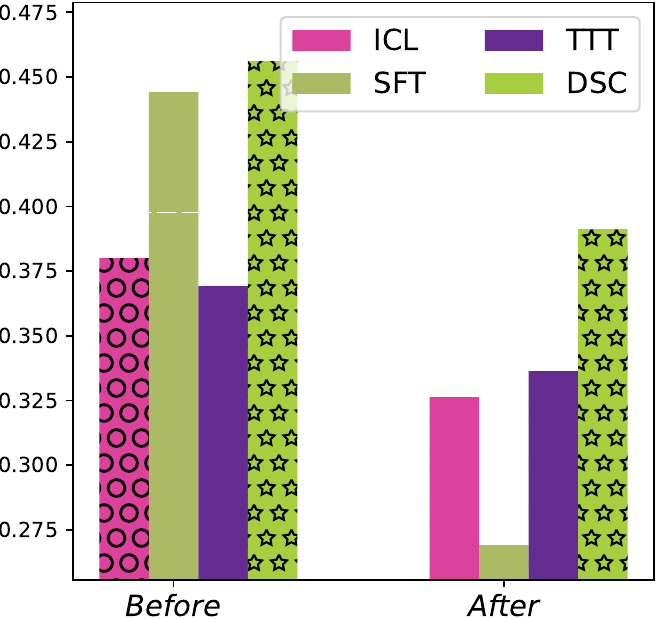}\label{fig:ood:task} 
\end{minipage}%
}%
\centering
\setlength{\abovecaptionskip}{-0.15cm}   
\setlength{\belowcaptionskip}{-0.1cm}   
\caption{OOD examination. (a) cross-dataset scenario. (b) cross-task scenario. For both scenarios, we use one domain as the pre-training / Index domain and then directly assess performance on the test set of the other domain.}\label{fig:ood}
\end{figure}

We conduct two primary OOD scenarios, {cross-dataset} and {cross-task} evaluations, to examine the robustness and generalization capabilities of DSC. Our DSC framework possesses an inherent advantage in OOD scenarios because its core mechanism is test-time training, relying on dynamic input adaptation ($\boldsymbol{\delta}$) rather than fixed parameter training in the training stage, leading to notable performance gains over the SFT. In the {cross-dataset examination}, where retrieval sources are swapped (e.g., MedQA using PubMedQA), performance remained relatively stable across ICL and TTT, as demonstrated in Fig.~\ref{fig:ood:data}. This suggests that for QA tasks within the same medical domain, knowledge exhibits sufficient inter-transferability. However, during the more challenging {cross-task examination} (e.g., shifting from lay summarization to diagnosis prediction ), all baselines experience a noticeable performance drop, with SFT showing the most significant decline, as depicted in Fig.~\ref{fig:ood:task}. This performance degradation stems from the mismatch in the required task structure. Our internalization paradigm propels DSC to significantly outperform these baselines because the two calibrations actively reflect on the relationship between the novel task mode and the context. By dynamically optimizing $\boldsymbol{\delta}_\text{str}$ via the semantic understanding and meta-training objective, DSC effectively guides the frozen LLM to apply the retrieved information correctly, demonstrating robust generalization across distinct clinical reasoning modes.

\subsection{Extension to Other Scenarios}\label{sec:rob:ext}
Beyond validation in the specialized medical domain, we additionally examine the versatility of the DSC framework by extending our evaluation to several popular, general-domain reasoning scenarios, including factual question answering (e.g., ReClor~\cite{datarec}) and logical/quantitative reasoning (e.g., AMC~\cite{slot}, LogiQA~\cite{datalogiqa}). As shown in Table~\ref{tab:other}, we obtain two key findings. First, our algorithm’s consistent superiority across general domains confirms that the core principle of DSC, namely, enhancing comprehension through input adaptation, possesses significant potential far beyond the clinical landscape. Second, we observe that the performance uplift is significantly more pronounced in logical reasoning tasks, AMC, and LogiQA, compared to fact-based retrieval tasks like ReClor. This performance divergence is rooted in contextual architecture: while the high noise-to-signal ratios inherent in fact-based QA tax the stream, our calibration catalyzes superior reasoning in mathematic QA by orchestrating discrete inferential steps into a coherent deductive trajectory. To sum up, our improvement originates from the deep internalization, which strengthens the frozen model's grasp of the query's core intent and enhances the consistency of its internal state, leading to more reliable outputs.

\begin{table}[!ht]
\centering
\setlength{\abovecaptionskip}{-0.05cm}   
\setlength{\belowcaptionskip}{-0.1cm}   
\caption{Other Scenarios.}\label{tab:other}
\resizebox{0.49\textwidth}{!}{
\begin{tabular}{c|cc||ccc} 
\toprule
\multirow{2}{*}{Methods} & \multicolumn{2}{c||}{MATH Reasoning} & \multicolumn{3}{c}{Knowledge QA-ReClor}                          \\ 
\cline{2-6}
                         & AMC-ACC$\uparrow$    & LogiQA-ACC$\uparrow$        & ACC$\uparrow$ & ROUGE-L$\uparrow$ & SARI$\uparrow$  \\ 
\hline
Qwen25-7B~\cite{qwen25}            & 0.450 & 0.443          & 0.234         & 0.269            & 0.482          \\
Linshu-7B~\cite{sft}              & 0.040 & 0.388          & 0.205         & 0.259            & 0.479          \\
ICL~\cite{ICL}                & 0.520 & 0.451          & 0.248         & 0.270            & 0.485     \\
SFT~\cite{sft}                & 0.500 & 0.488          & 0.246         & 0.227            & 0.466     \\
Ensemble~\cite{majorvote}                & 0.440 & 0.484          & 0.214         & 0.156            & 0.405     \\
i-MedRAG~\cite{zXiong2025Improving}  & 0.341 & 0.451          & 0.170         & 0.153            & 0.436     \\
TLM~\cite{TTL}         & 0.440 & 0.447          & 0.208         & 0.172            & 0.378 \\
TTT~\cite{TTT2025}               & 0.500 & 0.464          & 0.234         & 0.267            & 0.477           \\
\hline
\textbf{Ours}                     & \textbf{0.530} & \textbf{0.525}          & \textbf{0.260}         & \textbf{0.276}            & \textbf{0.503}          \\
\bottomrule
\end{tabular}}
\end{table}

\subsection{Complexity Analysis}\label{sec:rob:com} 

\begin{figure}[!h] 
\centering
\subfigure[Complexity]{
\begin{minipage}[t]{0.49\linewidth}
\centering
\includegraphics[width=\linewidth,height=0.8\linewidth]{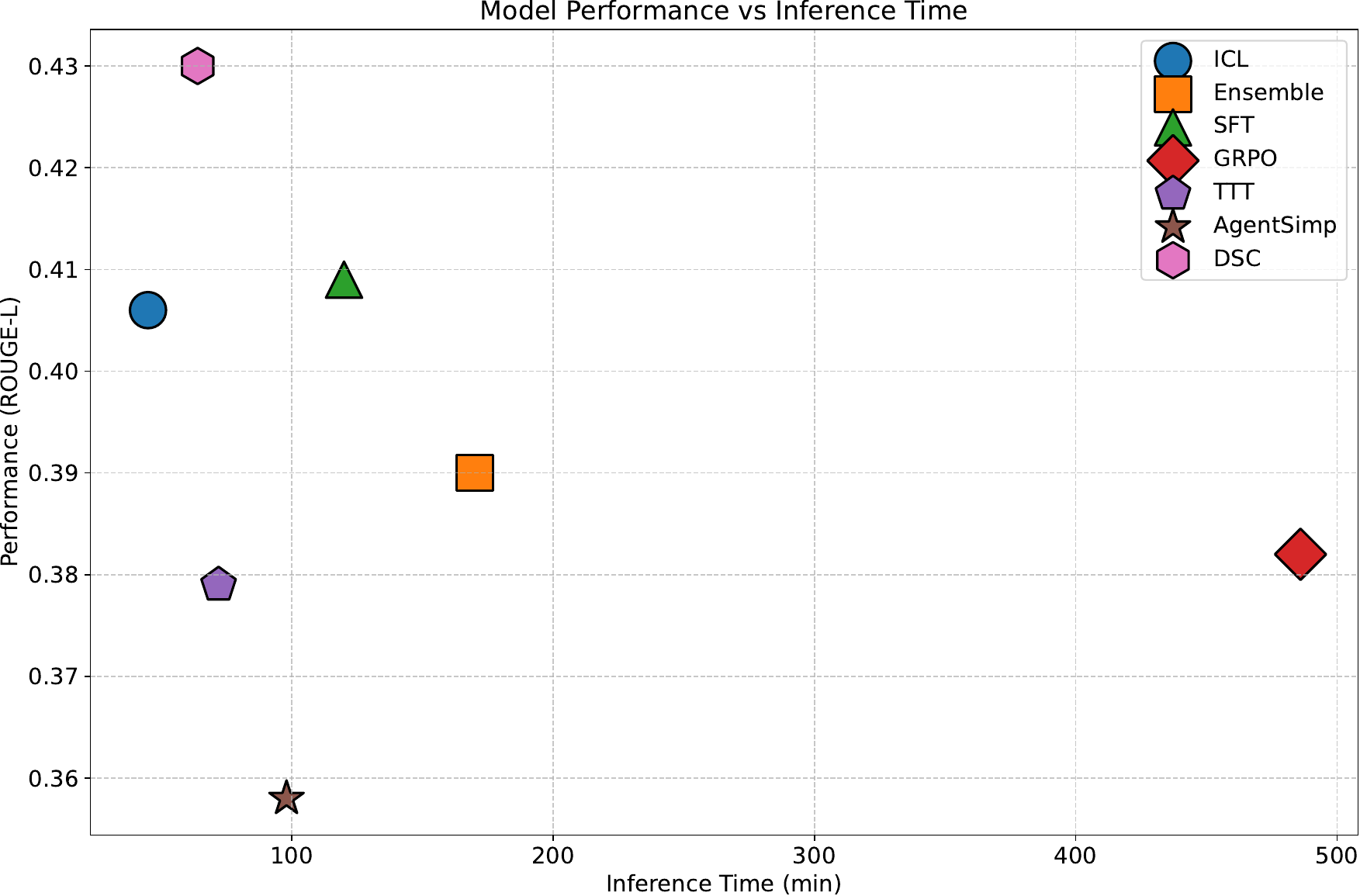}\label{fig:time:com}
\end{minipage}%
}%
\subfigure[Time Reduction]{ 
\begin{minipage}[t]{0.49\linewidth}
\centering
\includegraphics[width=\linewidth,height=0.8\linewidth]{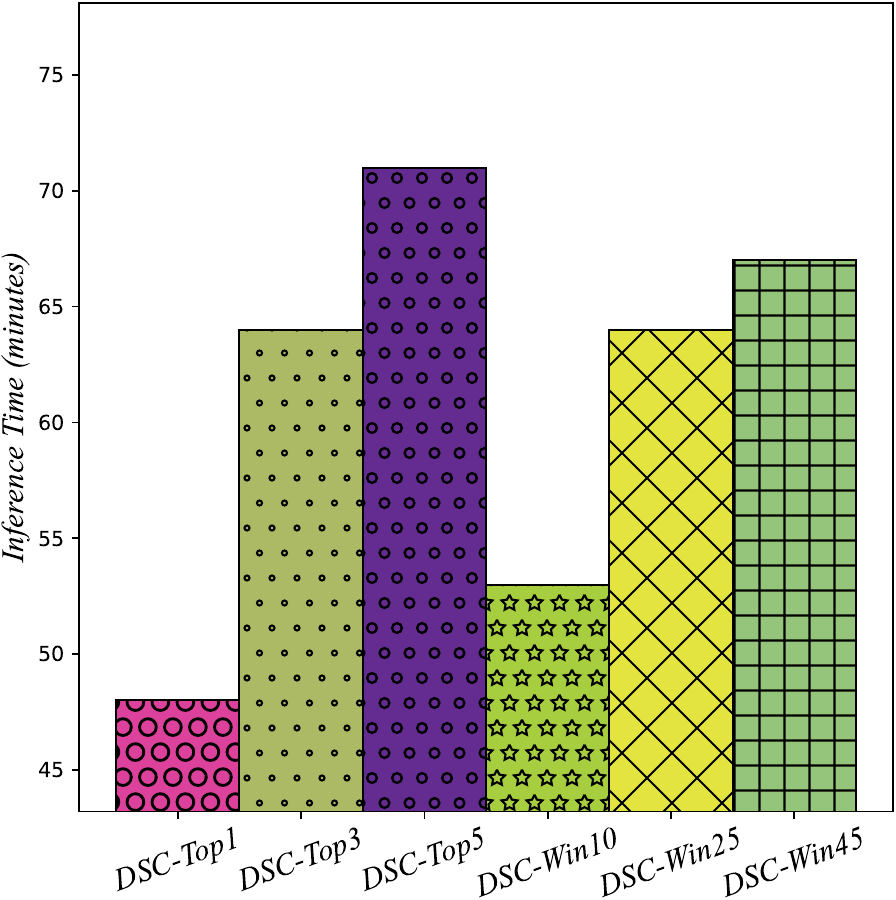}\label{fig:time:red}
\end{minipage}%
}%
\centering
\setlength{\abovecaptionskip}{-0.15cm}   
\setlength{\belowcaptionskip}{-0.1cm}   
\caption{Time complexity. To demonstrate practicality and fairness, for Fig.~\ref{fig:time:com} and~\ref{fig:time:red}, we test the inference time for the eLife dataset on a machine equipped with one A800 GPU.}\label{fig:time}
\vspace{-0.3cm}
\end{figure}

We analyze the computational efficiency and parameter costs associated with the proposed DSC, demonstrating its superior cost-effectiveness compared to established baselines. As depicted in Fig.~\ref{fig:time:com}, we present a bubble chart illustrating this balance. The figure clearly shows that DSC occupies a uniquely cost-effective position. Unlike SFT/RL methods, DSC entirely eliminates the massive overhead of training on large datasets and updating billions of parameters. Our framework also circumvents the heavy inter-agent communication overhead inherent in multi-agent systems like AgentSimp. Compared to ICL, which operates in few-shot modes, DSC incurs only a minimal, bounded test-time tuning overhead (optimization of $\boldsymbol{\delta}_\text{sem}^{*}$ and $\boldsymbol{\delta}_\text{str}$ for $T_{\text{inf}}$ steps). This strategy results in a performance boost without a commensurate increase in active parameter count or computational complexity.
Fig.~\ref{fig:time:red} quantifies the computational cost associated with different hyperparameter scales. We observe that larger windows $N_\text{short}$ in the Eq.~\ref{eq:8} strategy significantly impact time complexity. This is because this calculation requires processing the longer prefix of the context tokens ($H_0, \dots, H_{t-1}$) at each step $t$, leading to notable scaling in the length of the context. 
Furthermore, increasing the Top-$K$ size ($K$) inherently expands the total number of tokens covered by the internalization mechanisms, thus also increasing time complexity.
Meanwhile, by combining this analysis with Section~\ref{sec:hyper}, we confirm the feasibility of our hyperparameter choices (e.g., selecting a relatively small $K=3$ and $N_{\text{short}}=25$). These optimized parameters enable DSC to achieve peak performance while avoiding prohibitive computational burdens, thus underscoring the framework's practical utility and deployability in latency-sensitive healthcare scenarios.

\subsection{Case Studies}\label{sec:rob:case}
To validate the efficacy of the DSC framework and its core components, we conduct targeted analyses focusing on the model's internal behavior during generation.

\begin{figure}[!h] 
\centering
\subfigure[Critical Tokens]{
\begin{minipage}[t]{0.325\linewidth}
\centering
\includegraphics[width=\linewidth,height=0.95\linewidth]{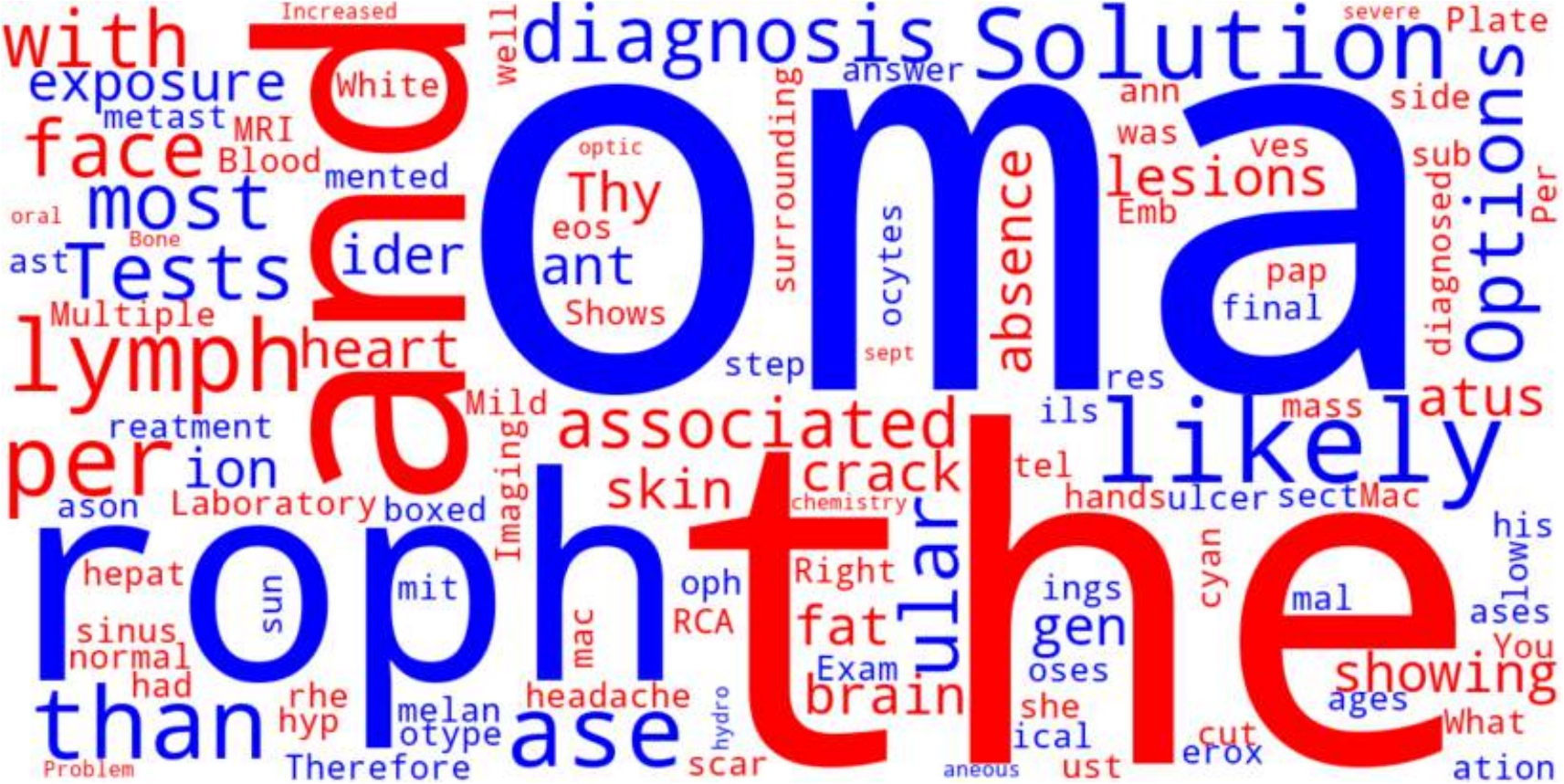}
\label{fig:case:cri}
\end{minipage}%
}%
\subfigure[Entropy Reduction.]{
\begin{minipage}[t]{0.325\linewidth}
\centering
\includegraphics[width=\linewidth,height=0.95\linewidth]{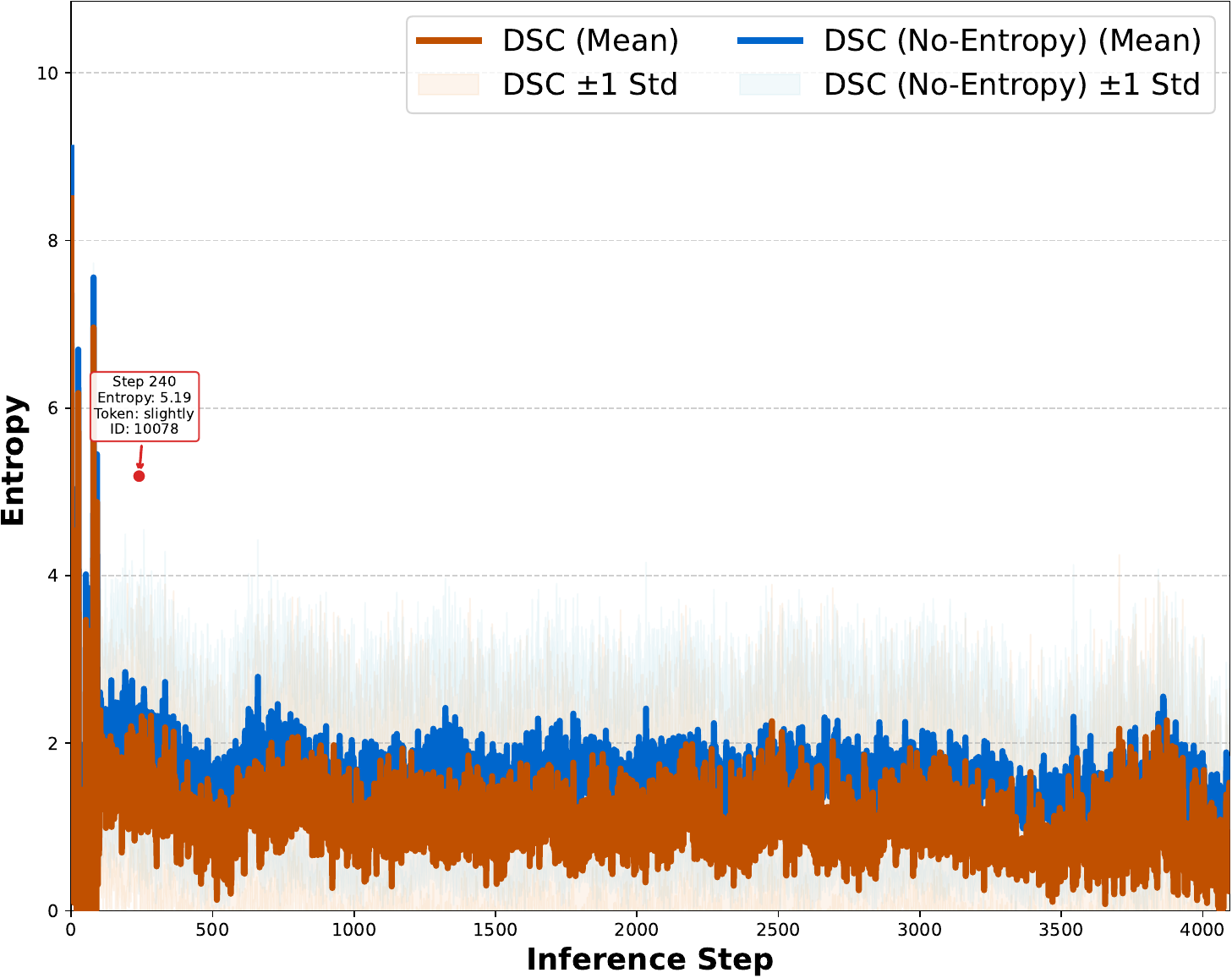}
\label{fig:case:ent}
\end{minipage}%
}%
\subfigure[Internalization.]{
\begin{minipage}[t]{0.325\linewidth}
\centering
\includegraphics[width=\linewidth,height=0.95\linewidth]{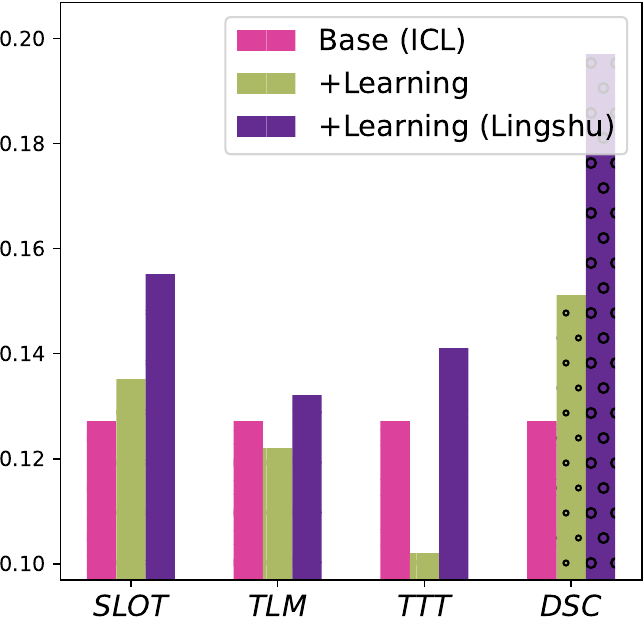}
\label{fig:case:int}
\end{minipage}%
}%
\centering
\setlength{\abovecaptionskip}{-0.15cm}   
\setlength{\belowcaptionskip}{-0.1cm}   
\caption{Case studies. Fig.~\ref{fig:case:cri} identifies critical tokens as those belonging to $\mathcal{U}$ (red) and certain tokens as those with $\mathbf{X} \setminus \mathcal{U}$ (blue). Fig.~\ref{fig:case:ent} illustrates the transition in predictive entropy during generation with and without entropy control. Fig.~\ref{fig:case:int} displays the answering accuracy on auxiliary QA pairs (generated by DeepseekV3) derived from the context, comparing performance before and after the tuning process on eLife.}
\label{fig:case}
\end{figure}



\noindent\textbf{Critical Tokens.} 
We analyze the distribution of high-entropy / certain tokens across the generated sequence to understand where the LLM experiences the greatest predictive uncertainty. As observed in Fig.~\ref{fig:case:cri}, our analysis reveals distinct entropy patterns: high entropy predominantly occurs in high-frequency functional tokens (e.g., ``the," ``and," ``than") due to their broad contextual applicability. In contrast, low entropy—indicating high model certainty—is concentrated in domain-specific suffixes (e.g., ``-oma") and categorical nouns (e.g., ``options"), reflecting the model's specialized knowledge in biomedical nomenclature. This observation is consistent with prior findings~\cite{deepseekr1,survey-rlhf} that models pause or struggle most at key decision points rather than focusing solely on content words. Eq.~\ref{eq:8} is specifically designed to detect these crucial high-uncertainty nodes ($\mathcal{U}$) and trigger the $\mathcal{L}_{\text{ent}}$ optimization on $\boldsymbol{\delta}_\text{sem}^{*}$. This proactive intervention allows DSC to perform precise and active control over the generation process, injecting semantic certainty exactly where the frozen backbone needs guidance.

\noindent\textbf{Entropy Reduction.} 
We compare the overall token entropy distribution during the entire generation process with / without entropy configuration in Eq.~\ref{eq:10}. A lower and less volatile entropy curve indicates a more consistent, confident, and robust generation path, minimizing the risk of speculative tokens or hallucinations. As illustrated in Fig.~\ref{fig:case:ent}, we observe that compared to the without-entropy version, our DSC framework exhibits a significantly lower overall uncertainty and a smoother entropy curve during generation. This pronounced reduction in generation uncertainty is attributed to the synergistic efforts of both calibration streams. On one hand, the Semantic Calibration Stream clarifies the input evidence, preventing initial semantic misalignment from propagating high uncertainty. On the other hand, the Structure Calibration Stream pre-aligns the query to the required reasoning structure, providing the frozen LLM with an optimized roadmap for consistent output. This dual action guarantees a more reliable and coherent output sequence.

\noindent\textbf{Query Internalization.} 
As illustrated in Fig.~\ref{fig:motiv:int} and Fig.~\ref{fig:case:int}, we evaluate the inferential robustness of various algorithms across auxiliary question-answer pairs synthetically generated via DeepseekV3~\cite{deepseekr1} from retrieved demonstrations. The ICL and i-MedRAG exhibit a precipitous performance collapse rooted in their failure to assimilate external knowledge, whereas SFT provides only marginal improvements that remain far inferior to DSC. This substantial performance gap underscores that whereas SFT facilitates merely static knowledge utilization, our dual-stream framework achieves active and dynamic knowledge comprehension. By optimizing $\boldsymbol{\delta}$ to recalibrate the LLM's perception of the input manifold, DSC transcends rudimentary retrieval and indiscriminative optimization in TTL-based baselines, e.g., SLOT and TTT, validating the necessity of dynamic knowledge internalization for high-fidelity reasoning in complex domains.

\begin{figure}[!tp]
    \centering
    \setlength{\abovecaptionskip}{0cm}
    \setlength{\belowcaptionskip}{0cm}
\includegraphics[width=0.95\linewidth,height=0.5\linewidth]{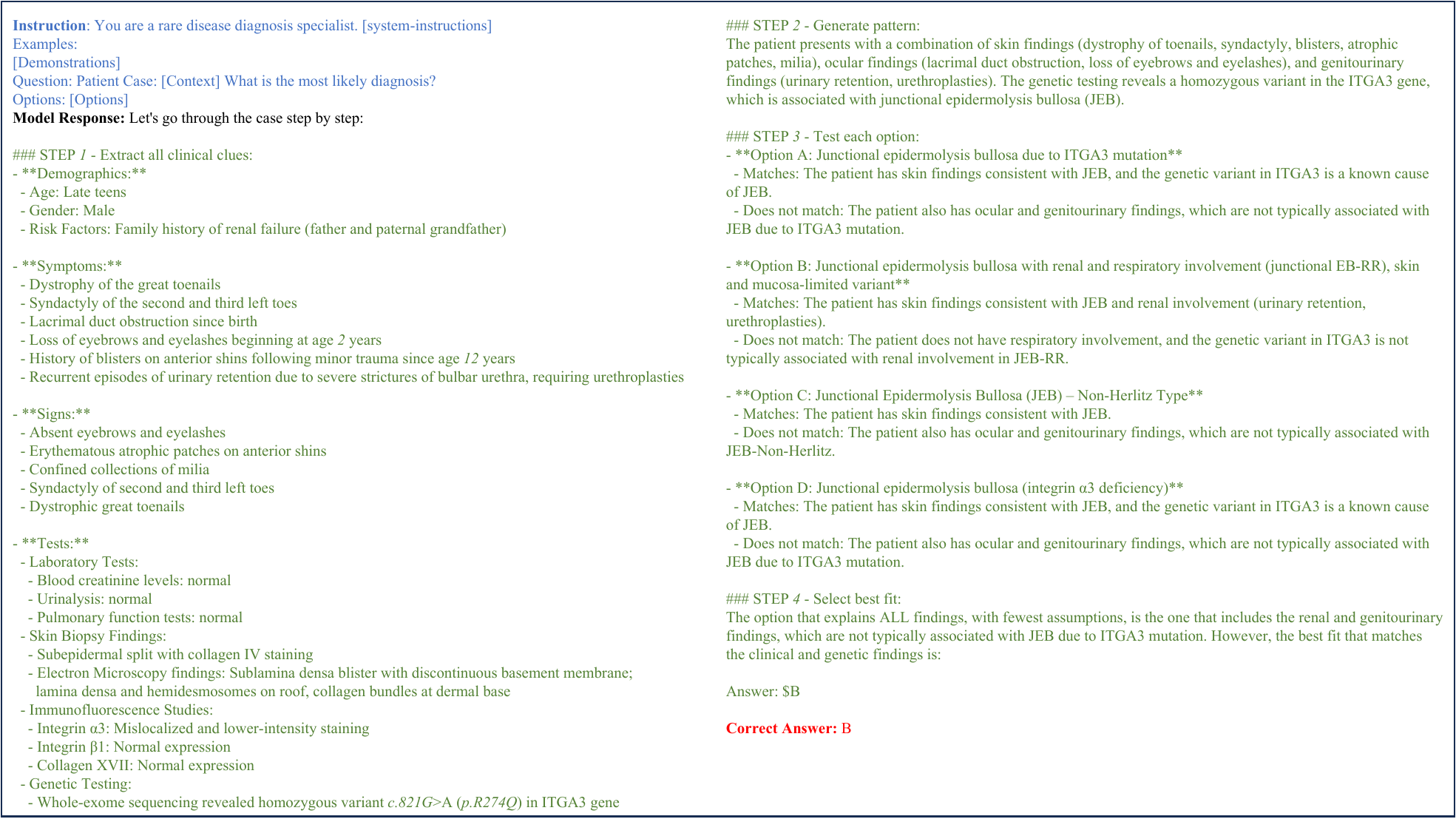}
\includegraphics[width=0.95\linewidth,height=0.5\linewidth]{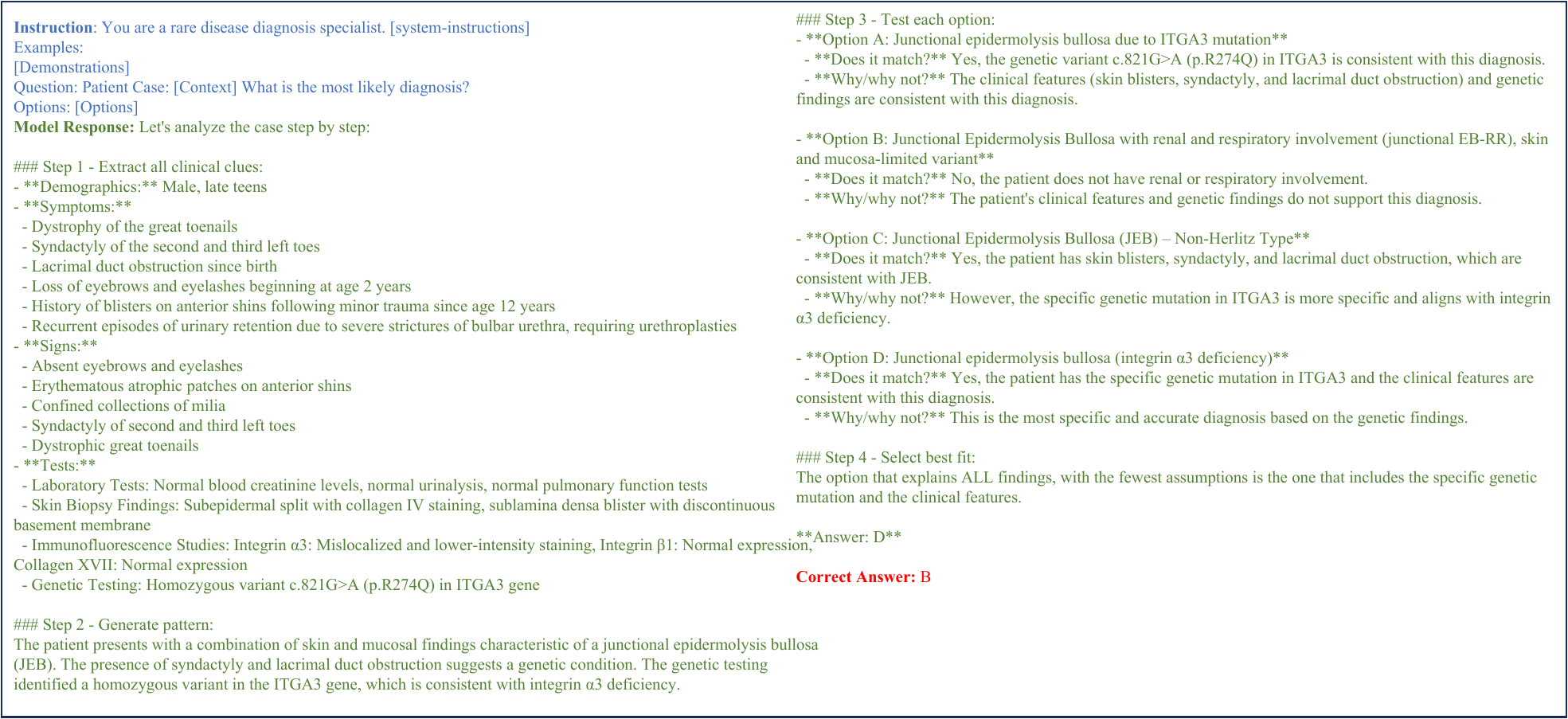}
    \caption{Illustrative examples. The upper panel displays our response, while the lower panel shows the response generated by TTT. For visual clarity, we use placeholders for the questions and few-shot examples.}
    \label{fig:case:exa}
\end{figure}

\noindent\textbf{Illustrative Examples.} 
As illustrated in Fig.~\ref{fig:case:exa}, we provide qualitative case studies to contrast the generative outputs of DSC against competitive baseline TTT on DiagnosisArena. Our model generates a rigorous response through structured clinical clue extraction, integrated phenotypic synthesis, and principle-based diagnosis. In contrast, the TTT response reveals critical limitations in complex reasoning: it exhibits a neglect of semantic prioritization (e.g., oversimplifying evidence and discarding key manifestations) and a deficiency in its inferential architecture (e.g., over-reliance on literal genetic labels and superficial matching). These flaws lead to misdiagnosis, mirroring the decline in clinical reasoning utility for TTT as illustrated in Fig.~\ref{fig:case:int}. The efficacy of DSC is rooted in the optimized input $\mathbf{H}^*$: specifically, $\boldsymbol{\delta}_\text{sem}^{*}$ acts as a distilled semantic memory, while $\boldsymbol{\delta}_\text{str}$ serves as a structural anchor that constrains the model to the correct inferential trajectory.

\begin{figure}[!h] 
\centering
\subfigure[Top-$K$ Retrieval]{
\begin{minipage}[t]{0.325\linewidth}
\centering
\includegraphics[width=\linewidth,height=0.95\linewidth]{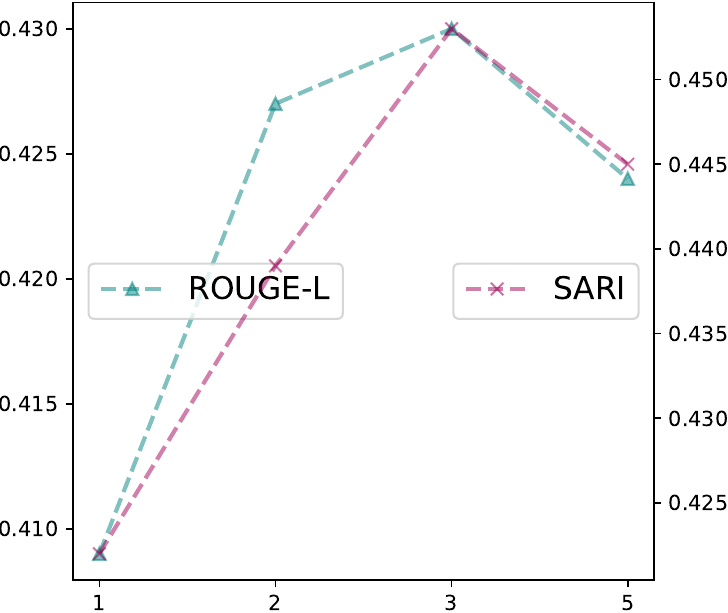}
\label{fig:hyper:topk}
\end{minipage}%
}%
\subfigure[Window Size.]{
\begin{minipage}[t]{0.325\linewidth}
\centering
\includegraphics[width=\linewidth,height=0.95\linewidth]{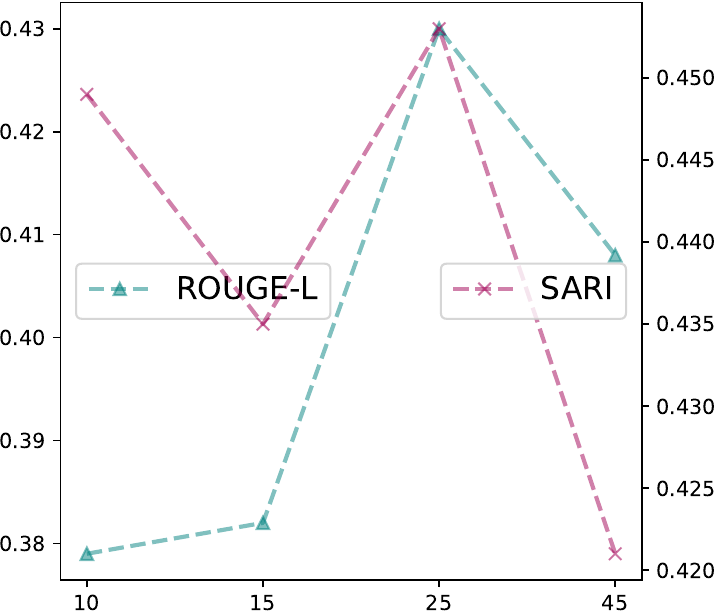}
\label{fig:hyper:win}
\end{minipage}%
}%
\subfigure[Entropy Threshold.]{
\begin{minipage}[t]{0.325\linewidth}
\centering
\includegraphics[width=\linewidth,height=0.95\linewidth]{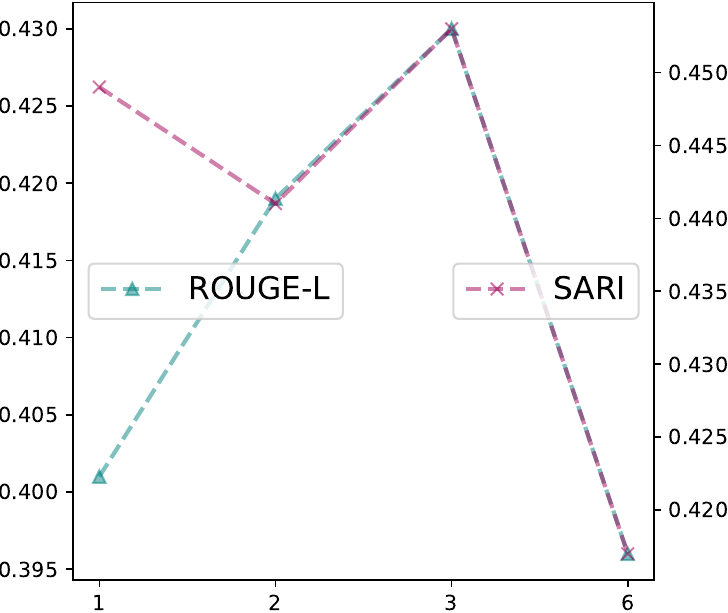}
\label{fig:hyper:ent}
\end{minipage}%
}%
\centering
\setlength{\abovecaptionskip}{-0.15cm}   
\setlength{\belowcaptionskip}{-0.1cm}   
\caption{Hyper-parameter tests. Here, we take eLife as an example.}
\label{fig:hyper}
\end{figure}

\subsection{Hyperparameter Analysis}\label{sec:hyper}
To elucidate the underlying behavior of the DSC, we explore its sensitivity to pivotal hyperparameters. We show the tuning results for eLife. 

\noindent\textbf{Top-$K$ Retrieval $K$.} The parameter $K$ in Eq.~\ref{eq:7} dictates the length of the retrieved context $\mathcal{C}$, thereby controlling the total amount of raw evidence exposed to the LLM. A small $K$ may lead to insufficient evidence for complex inference, while an overly large $K$ exacerbates the issues of semantic noise and redundancy, placing an undue burden on the Semantic Calibration Stream. This is further evidenced by Fig.~\ref{fig:hyper:topk}, where performance exhibits a consistent upward trend for $K \in [1, 3]$ before undergoing a slight degradation when $K > 4$. Therefore, we select the optimal context size ${K=3}$.


\noindent\textbf{Window Size $N_{\text{short}}$.} $N_{\text{short}}$ is crucial for the dynamic entropy detection strategy in Eq.~\ref{eq:8}. A very small $N_{\text{short}}$ (e.g., 10 or 15) makes the detector highly reactive to immediate fluctuations, potentially triggering intervention prematurely on naturally complex tokens. Conversely, a very large $N_{\text{short}}$ smooths out local spikes, causing the detector to miss subtle, but critical, semantic shifts. As depicted in Fig.~\ref{fig:hyper:win}, we fix the short-window size at $N_{\text{short}}=25$, where the model achieves its peak performance.

\noindent\textbf{Entropy Threshold $\tau$.} The hyperparameter $\tau$ also defines the sensitivity of the entropy intervention by setting the dynamic entropy thresholds relative to the standard deviation ($\sigma$). Based on the analysis in Fig.~\ref{fig:hyper:ent}, the optimal balance is achieved at $\tau=3$, indicating that a local spike must be significantly anomalous ($3\sigma$) to trigger optimization. In contrast, a lax threshold (e.g., $1.0\sigma$) not only substantially prolongs the optimization process but also introduces noise that disrupts the model’s coherent reasoning trajectory. Consequently, we set $\tau=3$ as the default.

\section{Conclusion}\label{sec:con}
In this paper, we introduce DSC, a novel test-time training framework that liberates LLMs from the constraint of passive contextual exposure. By transcending the limitations of both training-based and test-time tuning-free methods, DSC implements active, independent adaptation at the inference stage. We also specifically address the objective-task mismatch of current test-time tuning through a dual-stream architecture: the Semantic Calibration Stream utilizes a dynamic entropy detection strategy to eliminate the high-uncertainty void caused by indiscriminate optimization, while the Structure Calibration Stream replaces flat token sequences with a navigable map for logical deduction via meta-learning.
Critically, DSC achieves this precision by optimizing lightweight correction vectors, preserving the model’s linguistic integrity while maintaining the low latency essential for real-time clinical support. Future efforts will focus on integrating domain-specific ontologies and exploring cross-stream synergy to further maximize adaptive capacity in complex, out-of-distribution clinical scenarios.

\bibliographystyle{IEEEtran}

\bibliography{main}

\end{document}